\pdfoutput=1
\documentclass[11pt,aps,a4paper,eqsecnum,amsmath,amssymb
              ,nofootinbib
              ,notitlepage%,endfloats*
              ,longbibliography
              ]{revtex4-1}

\usepackage{graphicx}
\usepackage{pict2e}

\def\Wop[#1][#2][#3][#4][#5]{W
  \left(\left.\begin{array}{cc}
        #1&#2\\#3&#4\end{array}\right|\,#5
  \right)}
\def\Whop[#1][#2][#3][#4][#5]{W_h
  \left(\left.\begin{array}{@{}cc@{}}
        #1&#2\\#3&#4\end{array}\right|\,#5
  \right)}
  
\def\Wvop[#1][#2][#3][#4][#5]{W_v
  \left(\left.\begin{array}{@{}cc@{}}
        #1&#2\\#3&#4\end{array}\right|\,#5
  \right)}      
\def\Bop[#1][#2][#3][#4]{B
  \left(\left.\begin{array}{c}
        #1\\#3\end{array}\,#2\,\right|\,#4
  \right)}
  \def\BopP[#1][#2][#3][#4]{B'
  \left(\left.\begin{array}{c}
        #1\\#3\end{array}\,#2\,\right|\,#4
  \right)}
\def\BopR[#1][#2][#3][#4]{B
  \left(\left. #2 \, \begin{array}{c}
       #1\\#3\end{array}\,\right|\,#4
  \right)}
  \def\BopRP[#1][#2][#3][#4]{B'
  \left(\left. #2 \, \begin{array}{c}
       #1\\#3\end{array}\,\right|\,#4
  \right)}
\def\WopB[#1][#2][#3][#4][#5]{\bar{W}
  \left(\left.\begin{array}{cc}
        #1&#2\\#3&#4\end{array}\right|\,#5
  \right)}

\begin{document}
\preprint{} 

\title{Non-Abelian $SU(3)_k$ anyons: inversion identities for higher rank
  face models}

\author{Holger Frahm}

\author{Nikos Karaiskos} 
\altaffiliation[Present Address: ]{Max-Delbr\"uck-Center for
    Molecular Medicine, Berlin, Germany}

\affiliation{%
Institut f\"ur Theoretische Physik, Leibniz Universit\"at Hannover,
Appelstra\ss{}e 2, 30167 Hannover, Germany}
\date{August 26, 2015}

\begin{abstract}
  The spectral problem for an integrable system of particles satisfying the
  fusion rules of $SU(3)_k$ is expressed in terms of exact inversion
  identities satisfied by the commuting transfer matrices of the integrable
  fused $A_2^{(1)}$ interaction round a face (IRF) model of Jimbo, Miwa and
  Okado.  The identities are proven using local properties of the Boltzmann
  weights, in particular the Yang-Baxter equation and unitarity.  
  % They provide
  % a local separation of variables, similar as in the integrable vertex models
  % with the corresponding symmetries.
  They are closely related to the consistency conditions for the construction
  of eigenvalues obtained in the Separation of Variables approach to
  integrable vertex models.
\end{abstract}

\maketitle

%%%%%%%%%%%%%%%%%%%%%%%%%%%%%%%%%%%%%%%%%%%%%%%%%%%%%%%%%%%%%%%%%%%%%%
\section{Introduction}
Studies of integrable models \cite{Bethe31,Baxter:book,VladB,HUBBARD} in low
dimensions have provided important insights into the exotic properties of the
quasiparticle excitations in a correlated many-body system subject to strong
quantum fluctuations.  Particularly exotic objects are the non-Abelian anyons
with unconventional quantum statistics expected to be realized in certain
fractional quantum Hall states \cite{MoRe91,ReRe99}: interchanging these
quasi-particles can be described by a unitary rotation on the manifold of
robust degenerate states supported by a collection of a few of them.
These degeneracies will be lifted by interactions between the anyons and the
formation of collective states has been studied in models of interacting
anyons with given fusion and braiding properties
\cite{FTLT07,GATL09,GrSc09,BaSl09,GATH13,Finch.etal14}.

In this context, the local lattice Hamiltonians generated by the commuting
transfer matrix of integrable restricted solid on solid (RSOS) models
\cite{AnBF84} and their generalizations \cite{Pasq88,JiMO88a,FrZu90,Gepn92}
have recently attracted renewed interest, see e.g.\
\cite{FTLT07,FiFr13,FiFF14}.  In these lattice models the local state
variables (spins) take values from a given set of representations of a Lie
algebra $\mathfrak{g}$.  The possible pairs of spins on adjacent sites are
constrained by the fusion rules of the algebra \cite{Pasq88,FrZu90,Gepn92},
leading e.g.\ to the RSOS condition for $SU(2)_k$ anyons.  Below we consider
such models relevant to anyons satisfying higher rank fusion rules,
specifically of $SU(n)_k$ for $n=3$.  This fusion algebra is a truncation of
the category of irreducible representations (irreps) of the quantum group
$U_q[su(n)]$ with $q=\exp(2\pi i/(n+k))$ or, equivalently, the level-$k$
integrable representations of the Kac-Moody algebra
$A_{n-1}^{(1)}=\widehat{SU(n)}$.
Local Boltzmann weights for IRF models based on this algebra for anyons
corresponding to the fundamental vector representation which satisfy a
Yang-Baxter equation have been constructed by Jimbo, Miwa and Okado
\cite{JiMO88a}.  Using the fusion procedure the model has been further
generalized to allow for arbitrary $SU(n)_k$ anyons relating the admissible
spins on neighbouring sites in the horizontal and vertical direction,
respectively \cite{JKMO88}.  Based on the fusion hierarchy of functional
equations satisfied by the corresponding transfer matrices the central charge
and conformal weights of the conformal field theories describing the low
energy collective excitations of the model have been computed
\cite{BaRe90,ZhPe95}.

The goal of this paper is to provide a basis for a different approach towards
the solution of the spectral problem of this model based on a set exact
inversion identities satisfied by the transfer matrices (or their eigenvalues)
of inhomogeneous generalizations of these models.  For the six-vertex model
and the related RSOS models with integrable periodic or open boundary
conditions such identities have recently been derived from the underlying
Yang-Baxter or reflection equations for the local vertex weights using certain
physical assumptions such as crossing symmetry and unitarity
\cite{CYSW13a,FrKa14}, for the vertex models they also arise in Sklyanin's
separation of variables (SoV) approach \cite{Skly92}.
Complemented with information on the analytical properties of the transfer
matrix they can be related to the formulation of the spectral problem in the
form of Baxter's TQ-equation \cite{Baxt72a} or inhomogeneous generalizations
thereof which arise in models with non-diagonal boundary conditions breaking
the $U(1)$ bulk-symmetry of the vertex model \cite{CYSW13a,CaoX13,CYSW13}.
For the six-vertex model, the equivalence of these formulations can be shown
using the SoV approach where, in addition, the completeness of the spectrum
has been proven \cite{KiMN14}.

Just as in these cases the Boltzmann weights of the $SU(n)_k$ models
considered below satisfy a unitarity condition.  There is, however, no
crossing symmetry for the face weights which prevents the straightforward
extension of the results of Refs.~\cite{CYSW13a,FrKa14} to higher rank
symmetries.
Motivated by relations obtained from the fusion hierarchy of the integrable
$SU(3)$-invariant vertex model and the corresponding SoV
\cite{KuRe82,Skly93,PrSt00}, we find that a closed set of discrete inversion
identities can be formulated for \emph{two} transfer matrices from the fusion
hierachy of the IRF model corresponding to adjacency conditions in the
vertical direction induced by the anyon in the fundamental vector
representation and its dual, respectively.\footnote{%
  Related identities for $SU(n)$ vertex models with various boundary
  conditions have been constructed in Ref.~\cite{CYSW14a}.}

Our paper is organized as follows: to introduce our notation we first briefly
recall the definition of the $SU(n)_k$ anyon (or $A_{n-1}^{(1)}$ IRF) model
and its algebraic structure.  Then, motivated by the results for the $SU(3)$
vertex model \cite{KuRe82,KuRe83,Skly93} summarized in the appendix, we
introduce a generalized model based on fused Boltzmann weights and define the
transfer matrices appearing in the inversion identities for the $SU(3)_k$
model.  The main result of this paper is the proof of these identities in
Section~\ref{sec:proof}.  The paper ends with a brief discussion.

\section{$SU(n)_k$ anyon models}
Anyonic models can be decomposed into a finite set of topological sectors.
The corresponding conserved charges obey a commutative and associative fusion
algebra
\begin{equation}
  \label{fusalg}
  \psi_a\otimes\psi_b \cong \bigoplus_c N_{ab}^{c} \psi_c
\end{equation}
with non-negative integers $N_{ab}^{c}$.  In a graphical representaton (to be
read from top left to right) of fusion the vertex
\begin{equation*}
  \begin{picture}(50,40)(0,0)
    \put(5,5){\line(1,0){40}}
    \put(25,5){\line(-0.5,1){10}}
    \put(-5,2){$a$}
    \put(10,28){$b$}
    \put(50,2){$c$}
  \end{picture}
\end{equation*}
may occur provided that $N_{ab}^c\ne0$.  The fusion algebra places constraints
on the allowed sequence of charges in the basis of fusion path states for a
model of anyons $\psi_x$
\begin{equation}
\label{fpstate}
  |\cdots a_{n-1} a_n a_{n+1}\cdots\rangle = \ldots
  \begin{picture}(140,35)(0,0)
    \put(5,1){\line(1,0){130}}
    \put(40,1){\line(-0.5,1){10}}
    \put(80,1){\line(-0.5,1){10}}
    \put(120,1){\line(-0.5,1){10}}
    \put(10,-10){$a_{n-1}$}
    \put(55,-10){$a_{n}$}
    \put(90,-10){$a_{n+1}$}
    \put(20,28){$\psi_x$}
    \put(60,28){$\psi_x$}
    \put(100,28){$\psi_x$}
  \end{picture}
  \ldots
\end{equation}
i.e.\ $\psi_{a_{n+1}}$ has to appear in the fusion $\psi_{a_n}\otimes \psi_x$.

%%%%%%%%%%%%%%%%%%%%%%%%%%%%%%%%%%%%%%%%%%%%%%%%%%%%%%%%%%%%%%%%%%%%%%
\subsection{Local states and admissible pairs in the IRF mdel}
For the $SU(n)_k$ anyons considered in this paper the sectors are identified
with certain irreducible representations of the quantum group $U_q[su(n)]$ and
the fusion algebra is given by the decomposition of their tensor products into
irreps.  Just as the level $k$ dominant integral weights of $A_{n-1}^{(1)}$
the topological sectors in the $SU(n)_k$ anyon models are represented by
vectors
\begin{equation}
  \label{domwgt}
  a = \sum_{i=0}^{n-1} a_i\,\omega_i\,,\qquad \sum_{i=0}^{n-1} a_i = k\,
\end{equation}
with nonnegative integers $a_i$ and $\{\omega_i\}_{i=0}^{n-1}$ being the
fundamental weights of $A_{n-1}^{(1)}$ with $\omega_n=\omega_0$.  For given
$k\ge1$ the topological charges (or the spin variables in the IRF model) take
values in the set $P_+(n;k)$ of dominant weights (\ref{domwgt}).
A convenient representation of these local states is in terms of Young
diagrams $[\lambda_1,\ldots,\lambda_{n-1}]$ with $\lambda_i$ nodes in the
$i^{\mathrm{th}}$ row: for $k\equiv \lambda_0\ge \lambda_1\ge \ldots \ge
\lambda_{n-1}\ge \lambda_n\equiv 0$ this diagram can be identified with the
local state
\begin{equation} 
  \label{l_state}
  a = \sum_{i=0}^{n-1} \left(\lambda_i-\lambda_{i+1}\right)\,\omega_i 
  = \lambda_0\,\omega_0 + \sum_{i=1}^{n-1} \lambda_i\,e_i
  %\left(\omega_i-\omega_{i-1}\right)
  \,\in\, P_+(n;k)
\end{equation}
with the elementary vectors $e_i=\omega_i-\omega_{i-1}$.  Two diagrams
represent the same element if and only if one is obtained from the other by
removal of columns of height $n$.

As discussed above the fusion algebra (\ref{fusalg}) leads to a set of
constraints for an ordered pair of local states $(a,b)$ with $a,b\in P_+(n;k)$
to be admissible as a configuration on neighbouring sites of the lattice
model.
% for a system of anyons $\psi_\lambda$ corresponding to the dominant
% weight $\hat{\lambda}$ such a pair is allowed if $\psi_b$ appears in the
% decomposition of $\psi_a\otimes \psi_{\lambda}$.  
For a system of anyons $\psi_{[\lambda]}$ corresponding to the Young diagram
$[{\lambda}]$ such a pair is allowed if $N_{a[\lambda]}^{b}\ne 0$.
Considering anyons in the fundamental vector representation
$[1]\equiv[1,0,\ldots,0]$
the vertex
\begin{equation*}
  \begin{picture}(50,40)(0,0)
    \put(5,5){\line(1,0){40}}
    \put(25,5){\line(-0.5,1){10}}
    \put(-5,2){$a$}
    \put(10,28){$\psi_{[1]}$}
    \put(50,2){$b$}
  \end{picture}  
\end{equation*}
may occur in a fusion path state provided that
\begin{equation}
  \label{admissible_cond}
  b = a + e_i\,, \qquad \textrm{for some } i = 1, \ldots,n\,,  
\end{equation}
see e.g.\ Table~\ref{tab:fussu32} for $SU(3)_2$ anyons.
\begin{table}
\begin{ruledtabular}
\begin{tabular}{c|cccccc}
$\otimes$ & $[0,0]$ & $[1,0]$ & $[1,1]$ 
        & $[2,0]$ & $[2,1]$ & $[2,2]$ \\\hline
$[1,0]$ & $[1,0]$ & $[1,1] \oplus [2,0]$ & $[0,0] \oplus [2,1]$
        & $[2,1]$ & $[1,0] \oplus [2,2]$ & $[1,1]$
\\
$[1,1]$ & $[1,1]$ & $[0,0] \oplus [2,1]$ & $[1,0] \oplus [2,2]$
        & $[1,0]$ & $[1,1]\oplus[2,0]$ & $[2,1]$ \\
\end{tabular}
\end{ruledtabular}
\caption{\label{tab:fussu32}
  Fusion rules for $SU(3)_2$ anyons involving the fundamental anyons and their
  adjoints corresponding to the Young diagrams $[1,0]$ and $[1,1]$,
  respectively.}
\end{table}

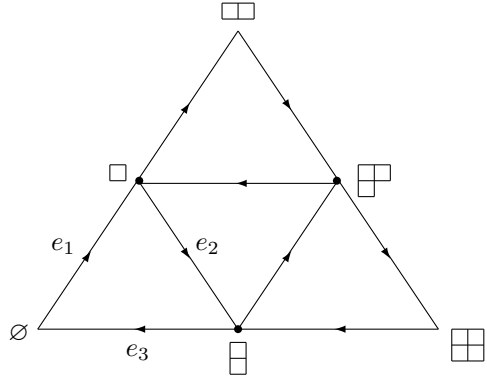
\begin{figure}[t]
\begin{picture}(150,150)(-0,-20)
\put(0,0){\line(2,3){75}}
\put(150,0){\line(-2,3){75}}
\put(0,0){\line(1,0){150}}
\put(75,0){\line(-2,3){37}}
\put(75,0){\line(2,3){37}}
\put(37,55){\line(1,0){75}}

\put(75,0){\circle*{3}}
\put(38,56){\circle*{3}}
\put(112,56){\circle*{3}}

\put(37,0){\vector(-1,0){1}}
\put(112,0){\vector(-1,0){1}}
\put(19,28){\vector(2,3){1}}
\put(56,84){\vector(2,3){1}}
\put(94,84){\vector(2,-3){1}}
\put(131,28){\vector(2,-3){1}}
\put(75,55){\vector(-1,0){1}}
\put(94,28){\vector(2,3){1}}
\put(56,28){\vector(2,-3){1}}

\put(-11,-4){$\varnothing$}

\put(27,56){\line(1,0){6}}
\put(27,56){\line(0,1){6}}
\put(27,62){\line(1,0){6}}
\put(33,56){\line(0,1){6}}

\put(69,123){\line(1,0){12}}
\put(69,123){\line(0,-1){6}}
\put(75,123){\line(0,-1){6}}
\put(81,123){\line(0,-1){6}}
\put(69,117){\line(1,0){12}}

\put(120,62){\line(1,0){12}}
\put(120,62){\line(0,-1){12}}
\put(126,62){\line(0,-1){12}}
\put(132,62){\line(0,-1){6}}
\put(120,56){\line(1,0){12}}
\put(120,50){\line(1,0){6}}

\put(155,0){\line(1,0){12}}
\put(155,0){\line(0,-1){12}}
\put(161,0){\line(0,-1){12}}
\put(167,0){\line(0,-1){12}}
\put(155,-6){\line(1,0){12}}
\put(155,-12){\line(1,0){12}}

\put(72,-5){\line(1,0){6}}
\put(72,-5){\line(0,-1){12}}
\put(78,-5){\line(0,-1){12}}
\put(72,-11){\line(1,0){6}}
\put(72,-17){\line(1,0){6}}

\put(33,-10){$e_3$}
\put(5,30){$e_1$}
\put(59,30){$e_2$}
\end{picture}
\caption{\label{fig:localweights} Adjacency graph for the set of local states
  $P_+(3; 2)$.  Admissible pairs of states in the anyon model corresponding to
  the fundamental vector representation $[1]$ are denoted by an arrow
  connecting the corresponding vertices.  For anyons $\psi_{[1^2]}$ in the
  adjoint representation the arrows would have to be reversed.}
\end{figure}

The set of local states and the constraints for a given type of anyons can be
represented in an oriented graph with nodes labeled by elements of $P_+(n;k)$
and arrows from node $a$ to $b$ representing an admissible pair $(a,b)$.  The
underlying fusion rules for $\psi_{[\lambda]}$ anyons are encoded in the
corresponding adjacency matrix $A^{[\lambda]}$ with elements
\begin{equation}
  \label{adjmat}
  \left(A^{[\lambda]}\right)_{ab} = N_{a[\lambda]}^b\,.
\end{equation}
Note that for $[\lambda]$ being a symmetric tensor or antisymmetric tensor the
decomposition $\psi_a\otimes\psi_{[\lambda]}$ is multiplicity free and the
elements of the adjacency matrix are $0$ or $1$.
The adjacency graph for the local states $P_+(3;2)$ in a system of
$\psi_{[1]}$ anyons is shown in Figure~\ref{fig:localweights}.

\subsection{Boltzmann weights}
%%%%%%%%%%%%%%%%%%%%%%%%%%%%%%%%%%%%%%%%%%%%%%%%%%%%%%%%%%%%%%%%%%%%%%
The $A_{n-1}^{(1)}$ IRF model for $SU(n)_k$ anyons in the vector
representation $[1]$ is defined on a square lattice, such that the spins on
the corners of a face take values in $P_+(n;k)$.  A Boltzmann weight
corresponding to a configuration $a,b,c,d$ round a face is depicted as
\begin{equation*}
  \Wop[c][d][b][a][u] =
  \begin{picture}(240,0)(230,12)
    % \put(130,11){$\Wop[c][d][b][a][u]$~~~~= }
    \put(250,0){\line(0,1){30}}
    \put(250,0){\line(1,0){30}}
    \put(280,0){\line(0,1){30}}
    \put(250,30){\line(1,0){30}}
    \put(265,0){\vector(-1,0){1}}
    \put(265,30){\vector(-1,0){1}}
    \put(280,15){\vector(0,1){1}}
    \put(250,15){\vector(0,1){1}}
    \put(262,12){$u$}
    \put(241,33){$c$}
    \put(282,33){$d$}
    \put(241,-9){$b$}
    \put(282,-9){$a$}
  \end{picture}
\end{equation*}
%\noindent
It is non-vanishing if and only if the four pairs of variables $(a,b)$,
$(b,c)$, $(d,c)$, $(a,d)$ are all admissible in the sense of
(\ref{admissible_cond}).
Their explicit expressions read  \cite{JiMO88a}
\begin{equation}
\begin{split}
\Wop[a + 2e_i][a+e_i][a+e_i][a][u] & = r_1(u)[u+1]\\
\Wop[a + e_i + e_j][a + e_i][a+e_i][a][u] & = r_1(u)
\frac{[u + a_{ji}][1]}{[a_{ji}]} \, , \qquad 
 \textrm{for} ~~ i \neq j  \\
\Wop[a + e_i + e_j][a+e_j][a+e_i][a][u] & = r_1(u)
 \frac{[u][a_{ji}-1]}{[a_{ji}]} \, , \qquad 
 \textrm{for} ~~ i \neq j \, .
 \label{Boltzmann_weights}
\end{split}
\end{equation}
For a state $a\in P_+(n;k)$ given by Eq.~(\ref{l_state}), $a_{ij}$ is defined
as the inner product
\begin{equation}
  a_{ij} = \langle a + \rho, e_i - e_j \rangle = 
  j-i+\lambda_i-\lambda_j \, ,
\end{equation}
with $\rho$ being the sum of fundamental weights.  At criticality, the
dependence of the Boltzmann weights on the spectral parameter $u$ is given by
trigonometric functions
\begin{equation}
[u] = \sin (u\eta)\,,\qquad \eta = \frac{\pi}{n+k} \, .
\end{equation}
These weights satisfy the initial condition
\begin{equation} 
\Wop[c][d][b][a][0] \sim \delta_{bd} \, .
\end{equation}
Notice also that at the special value $u=-1$ the weights containing straight
paths, that is from $a \to a + 2\,e_i$, are automatically excluded.  The
Boltzmann weights (\ref{Boltzmann_weights}) satisfy the Yang-Baxter equation
(YBE)
\begin{equation}
  \label{YBE}
  \begin{split}
    & \sum_g
    \Wop[a][g][f][e][v] \Wop[a][b][g][c][u] \Wop[g][c][e][d][u+v]\\
    &  \qquad =\sum_g
    \Wop[a][b][f][g][u+v] \Wop[f][g][e][d][u] \Wop[b][c][g][d][v] \, ,
  \end{split}
\end{equation}
and an inversion relation which can be used to fix the normalization 
function $r_1(u)$
\begin{equation}
  \sum_g 
  \Wop[d][g][a][b][u] \Wop[d][c][g][b][-u] = \delta_{ac} \, 
  r_1^2(u) [1-u][1+u] \, .
\end{equation}
There are additional symmetries, e.g.\ symmetrizability, $Z_n$-invariance and
a duality, see \cite{JiMO88a} for details.

For the proof of the inversion identities below we use graphical
representation of these local relations.  In particular, the YBE may be
depicted as
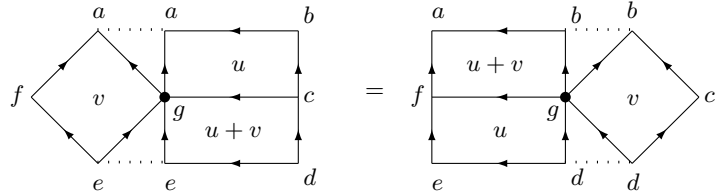
\begin{equation*}
\begin{picture}(400,80)(0,-10)
% LHS
\footnotesize
\put(100,0){\line(-1,1){25}}
\put(100,0){\line(1,1){25}}
\put(75,25){\line(1,1){25}}
\put(100,50){\line(1,-1){25}}
\put(98,22){$v$}
\put(67,23){$f$}
\put(98,55){$a$}
\put(98,-10){$e$}

\put(125,25){\circle*{4}}
\put(125,25){\line(0,1){25}}
\put(125,25){\line(0,-1){25}}
\put(125,50){\line(1,0){50}}
\put(125,25){\line(1,0){50}}
\put(125,0){\line(1,0){50}}
\put(175,25){\line(0,1){25}}
\put(175,25){\line(0,-1){25}}
\put(150,35){$u$}
\put(140,10){$u+v$}

\multiput(100,0)(4,0){7}{\line(0,1){1}}
\multiput(100,50)(4,0){7}{\line(0,1){1}}

\put(128,18){$g$}

\put(125,55){$a$}
\put(125,-10){$e$}
\put(177,53){$b$}
\put(177,23){$c$}
\put(177,-8){$d$}

\put(150,0){\vector(-1,0){1}}
\put(150,25){\vector(-1,0){1}}
\put(150,50){\vector(-1,0){1}}
\put(175,12){\vector(0,1){1}}
\put(125,12){\vector(0,1){1}}
\put(175,37){\vector(0,1){1}}
\put(125,37){\vector(0,1){1}}

\put(112,12){\vector(1,1){1}}
\put(88,12){\vector(-1,1){1}}
\put(112,38){\vector(-1,1){1}}
\put(88,38){\vector(1,1){1}}

\put(200,25){$=$}

% RHS
\put(225,25){\line(0,1){25}}
\put(225,25){\line(0,-1){25}}
\put(225,50){\line(1,0){50}}
\put(225,25){\line(1,0){50}}
\put(225,0){\line(1,0){50}}
\put(275,25){\line(0,1){25}}
\put(275,25){\line(0,-1){25}}
\put(238,35){$u+v$}
\put(248,10){$u$}

\put(225,55){$a$}
\put(225,-10){$e$}
\put(277,53){$b$}
\put(277,-8){$d$}
\put(217,23){$f$}

\put(300,0){\line(-1,1){25}}
\put(300,0){\line(1,1){25}}
\put(275,25){\line(1,1){25}}
\put(300,50){\line(1,-1){25}}
\put(275,25){\circle*{4}}
\put(298,22){$v$}
\put(268,18){$g$}
\put(298,55){$b$}
\put(298,-10){$d$}
\put(327,23){$c$}

\multiput(275,0)(4,0){7}{\line(0,1){1}}
\multiput(275,50)(4,0){7}{\line(0,1){1}}

\put(250,0){\vector(-1,0){1}}
\put(250,25){\vector(-1,0){1}}
\put(250,50){\vector(-1,0){1}}
\put(275,12){\vector(0,1){1}}
\put(225,12){\vector(0,1){1}}
\put(275,37){\vector(0,1){1}}
\put(225,37){\vector(0,1){1}}

\put(312,12){\vector(1,1){1}}
\put(288,12){\vector(-1,1){1}}
\put(312,38){\vector(-1,1){1}}
\put(288,38){\vector(1,1){1}}

\end{picture}
\end{equation*}
In this and in the following figures, nodes marked with a dot ($\bullet$)
represent spin variables which are summed over all possible local states and
nodes with equal spins are connected with a dotted line. Unitarity condition
is depicted in a similar fashion as
\begin{equation*}
\begin{picture}(90,55)(50,42)
\footnotesize
\put(50,20){\line(-1,1){25}}
\put(50,20){\line(1,1){25}}
\put(25,45){\line(1,1){25}}
\put(50,70){\line(1,-1){25}}
\put(47,42){$u$}
\put(17,43){$a$}
\put(48,75){$d$}
\put(48,10){$b$}

\multiput(50,20)(4,0){13}{\line(0,1){1}}
\multiput(50,70)(4,0){13}{\line(0,1){1}}

\put(75,45){\circle*{4}}

\put(100,20){\line(-1,1){25}}
\put(100,20){\line(1,1){25}}
\put(75,45){\line(1,1){25}}
\put(100,70){\line(1,-1){25}}
\put(92,42){$-u$}
\put(72,37){$g$}
\put(98,75){$d$}
\put(98,10){$b$}
\put(127,43){$c$}

\put(37.5,57.5){\vector(1,1){1}}
\put(37.5,32.5){\vector(-1,1){1}}
\put(62.5,57.5){\vector(-1,1){1}}
\put(62.5,32.5){\vector(1,1){1}}

\put(87.5,57.5){\vector(1,1){1}}
\put(87.5,32.5){\vector(-1,1){1}}
\put(112.5,57.5){\vector(-1,1){1}}
\put(112.5,32.5){\vector(1,1){1}}

\normalsize

% \put(143,25){$=$}
% \put(160,25){$\delta_{ac} \, 
% r_1^2(u) [1-u][1+u] \, .$}

\end{picture}
= \delta_{ac} r_1^2(u) [1-u][1+u] \, .
\end{equation*}
\vspace{2mm}

\subsection{Hecke algebra and projectors}
\label{sec:Hecke}
The algebraic structure underlying the $A_{n-1}^{(1)}$ RSOS models is
connected with a quotient of the braid group, namely the Hecke algebra
\cite{Jimb86a,Pasq88a,BaRe90}.  This connection leads to the construction of
projection operators, which are used later in the text in order to prove the
inversion identities satisfied by the transfer matrices.

Let first $|a_0 \cdots a_{L+1}\rangle$ be a fusion path state (\ref{fpstate})
for the anyons considerered, i.e.\ a sequence of local states $a_i$ such that
each pair $(a_i,a_{i+1})$ is admissible.  We define the Yang-Baxter operators
by their action on these states as
\begin{equation}
  \label{YBop}
  W_i(u) |a_0 \cdots a_{L+1} \rangle = 
  \sum_{b_i\in P_+(n;k)} \Wop[a_{i-1}][a_i][b_i][a_{i+1}][u] 
  |a_0 \cdots a_{i-1}b_i a_{i+1}\cdots a_{L+1} \rangle  \,
\end{equation}
for $i=1,\ldots,L$.
In terms of the Yang-Baxter operators, the YBE (\ref{YBE}) takes the 
form 
\begin{equation} 
  W_i(u) \, W_{i+1}(u+v) \, W_{i}(v)
  = W_{i+1}(v) \, W_{i}(u+v) \, W_{i+1}(u)\, , 
\end{equation}
which is reminiscent of the braid relation satisfied by the generators of the
braid group.  In fact, by proper choice of the normalization $r_1(u)$ of the
Boltzmann weights (\ref{Boltzmann_weights}) these operators can be written as
\begin{equation} 
  W_i(u) = e^{-i\eta u} \, \mathbb{I} 
         + e^{-i\eta} \, \frac{[u]}{[1]} \, X_i \, .
\end{equation}
Here the $X_i$'s are independent of the spectral parameter and satisfy
\begin{equation} 
  \begin{split}
    & X_i \, X_{i+1} \, X_i = X_{i+1} \, X_i \, X_{i+1} \cr
    & X_i^2 - (q-1)X_i - q  = 0 \cr 
    & [X_i, X_j] = 0 \, , \qquad |i-j| > 1 \, 
  \end{split}
\end{equation}
with $q=\exp(2i\eta)=\exp(2\pi i/(n+k))$ being the deformation parameter of
the underlying quantum group $U_q[su(n)]$.  This set of relations gives rise
to a representation of the Hecke algebra $H_{L+1}(q)$.  Note that in the $n=2$
case, that is in the RSOS models, the Hecke algebra truncates essentially to
the Temperley-Lieb algebra.
As was pointed out in \cite{BaRe90} one actually obtains the representation of
the quotient of the Hecke algebra in which $q$-analogues of the full Young
(anti-)symmetrizers $(P_\ell^\pm)^2=P_\ell^\pm$ for the $SU(n)_k$ model can be
defined.  They act on (admissible) sequences of local states of length
$\ell+2$ and satisfy
\begin{equation} 
  \begin{aligned}
    &P^-_{n}(i, \ldots, i+n+1) = 0 \, , \qquad \forall i \\
    & P^{+}_\ell (i, \ldots, i+\ell+1) = 0 \, , \qquad \forall i\, , 
    \quad \ell = k+1, \ldots , n+k-1 \, .
  \end{aligned}
\end{equation}
% 
% Note that $(P^\pm_\ell)^2 = P_\ell^\pm $ are projection operators which we
% will use below to prove the inversion identities satisfied by the transfer
% matrices of the model.  
In terms of the Yang-Baxter operators they can be
formally written as
\begin{equation} 
  \label{hecke-proj}
  \begin{split}
    & P_{\ell}^\pm(0,\ldots,\ell+1)  = \prod_{m=0}^{\ell-1} \left(
      \prod_{j=1}^{\ell-m} \frac{[1]}{[j+1]} W_j(\pm j)
    \right) \, .
  \end{split}
\end{equation}
%
% By this definition, the projector $P_\ell^\pm$ act on (admissible) sequences
% of local states of length $\ell+2$.
As an example, we have the following schematic representation for $P_{1,2}^-$
(as before the spin variables $c_j$ on dotted nodes are being summed over)
%
%\vskip1.5cm
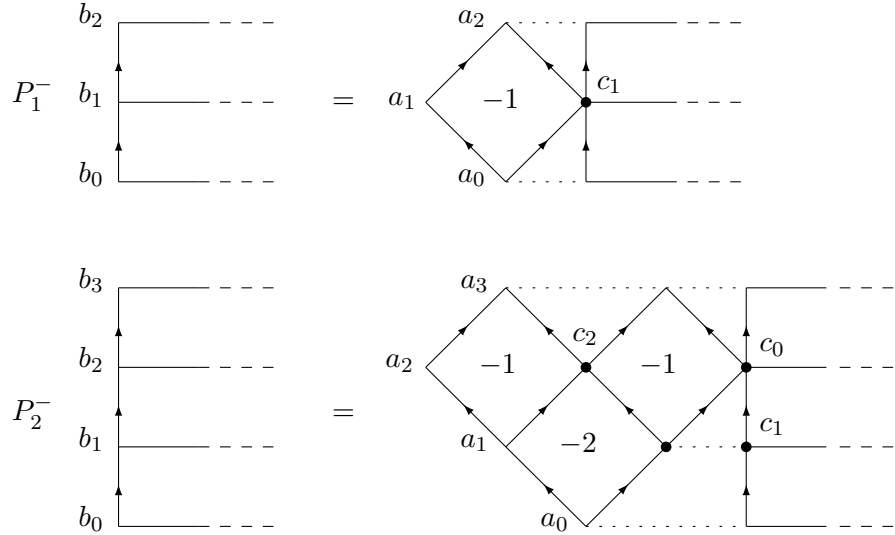
\begin{equation*}
\begin{picture}(310,200)(0,-160)
\put(-5,0){$P_1^-$}
\put(35,-30){\line(0,1){60}}
\put(35,30){\line(1,0){30}}
\put(35,0){\line(1,0){30}}
\put(35,-30){\line(1,0){30}}
\put(20,30){$b_2$}
\put(20,0){$b_1$}
\put(20,-30){$b_0$}
\multiput(65,30)(8,0){4}{\line(1,0){4}}
\multiput(65,0)(8,0){4}{\line(1,0){4}}
\multiput(65,-30)(8,0){4}{\line(1,0){4}}
\put(35,-15){\vector(0,1){1}}
\put(35,15){\vector(0,1){1}}
\put(115,-2){$=$}
\put(150,0){\line(1,1){30}}
\put(150,0){\line(1,-1){30}}
\put(180,30){\line(1,-1){30}}
\put(180,-30){\line(1,1){30}}
\put(210,0){\circle*{4}}
\put(171,-2){$-1$}
\put(210,30){\line(0,-1){60}}
\put(210,30){\line(1,0){30}}
\put(210,0){\line(1,0){30}}
\put(210,-30){\line(1,0){30}}
\multiput(240,30)(8,0){4}{\line(1,0){4}}
\multiput(240,0)(8,0){4}{\line(1,0){4}}
\multiput(240,-30)(8,0){4}{\line(1,0){4}}
\multiput(180,30)(5,0){6}{\line(1,0){1}}
\multiput(180,-30)(5,0){6}{\line(1,0){1}}
\put(162,30){$a_2$}
\put(136,-1){$a_1$}
\put(162,-30){$a_0$}
\put(214,5){$c_1$}
\put(210,-15){\vector(0,1){1}}
\put(210,15){\vector(0,1){1}}
\put(195,-15){\vector(1,1){1}}
\put(165,-15){\vector(-1,1){1}}
\put(195,15){\vector(-1,1){1}}
\put(165,15){\vector(1,1){1}}
%
% The second projector 
%
\put(-5,-120){$P_2^-$}
\put(35,-160){\line(0,1){90}}
\put(35,-70){\line(1,0){30}}
\put(35,-100){\line(1,0){30}}
\put(35,-130){\line(1,0){30}}
\put(35,-160){\line(1,0){30}}
\put(20,-70){$b_3$}
\put(20,-100){$b_2$}
\put(20,-130){$b_1$}
\put(20,-160){$b_0$}
\multiput(65,-70)(8,0){4}{\line(1,0){4}}
\multiput(65,-100)(8,0){4}{\line(1,0){4}}
\multiput(65,-130)(8,0){4}{\line(1,0){4}}
\multiput(65,-160)(8,0){4}{\line(1,0){4}}
\put(35,-145){\vector(0,1){1}}
\put(35,-115){\vector(0,1){1}}
\put(35,-85){\vector(0,1){1}}
\put(115,-120){$=$}
\put(150,-100){\line(1,1){30}}
\put(150,-100){\line(1,-1){60}}
\put(180,-70){\line(1,-1){60}}
\put(180,-130){\line(1,1){60}}
\put(210,-160){\line(1,1){60}}
\put(240,-70){\line(1,-1){30}}
\put(210,-100){\circle*{4}}
\put(270,-100){\circle*{4}}
\put(240,-130){\circle*{4}}
\put(270,-130){\circle*{4}}
\put(170,-102){$-1$}
\put(230,-102){$-1$}
\put(200,-132){$-2$}
\multiput(180,-70)(5,0){12}{\line(1,0){1}}
\multiput(240,-70)(5,0){6}{\line(1,0){1}}
\multiput(240,-130)(5,0){6}{\line(1,0){1}}
\multiput(210,-160)(5,0){12}{\line(1,0){1}}
\put(270,-160){\line(0,1){90}}
\put(270,-70){\line(1,0){30}}
\put(270,-100){\line(1,0){30}}
\put(270,-130){\line(1,0){30}}
\put(270,-160){\line(1,0){30}}
\put(163,-70){$a_3$}
\put(135,-100){$a_2$}
\put(163,-130){$a_1$}
\put(193,-160){$a_0$}
\put(205,-90){$c_2$}
\put(275,-94){$c_0$}
\put(275,-124){$c_1$}
\multiput(305,-70)(8,0){3}{\line(1,0){4}}
\multiput(305,-100)(8,0){3}{\line(1,0){4}}
\multiput(305,-130)(8,0){3}{\line(1,0){4}}
\multiput(305,-160)(8,0){3}{\line(1,0){4}}
\put(270,-145){\vector(0,1){1}}
\put(270,-115){\vector(0,1){1}}
\put(270,-85){\vector(0,1){1}}
\put(255,-115){\vector(1,1){1}}
\put(225,-145){\vector(1,1){1}}
\put(195,-115){\vector(1,1){1}}
\put(165,-85){\vector(1,1){1}}
\put(225,-85){\vector(1,1){1}}
\put(225,-115){\vector(-1,1){1}}
\put(165,-115){\vector(-1,1){1}}
\put(195,-145){\vector(-1,1){1}}
\put(195,-85){\vector(-1,1){1}}
\put(255,-85){\vector(-1,1){1}}
\end{picture}
\end{equation*}
The action of the projectors can be deduced from the explicit expressions of
the Boltzmann weights (\ref{Boltzmann_weights}).  First, one may observe that
for $u=-1$, the first of the weights (\ref{Boltzmann_weights}) is always zero.
Hence the operator $P_1^-$ projects out all states $(a_2,a_0)$ which are
connected by a straight line in the adjacency graph of local states
$P_+(n;k)$, see Figure~\ref{fig:localweights}.
Similarly, it can be shown that $P_{n-1}^- |a_0\cdots a_n\rangle \propto
\delta_{a_n a_0}$ for any fusion path state $|a_0\cdots a_n\rangle$ of
$SU(n)_k$ anyons.

% For the model that we focus on this action projects out
% all corner-to-corner paths. The action of the operator $P_2^-$ appears more
% complicated, but can be proven that it is proportional to
% $\delta_{a_3a_0}$. The exact coefficient of proportionality will be determined
% later by using the fused weights.

\section{Fused $A_{n-1}^{(1)}$ IRF models}
\subsection{Commuting transfer matrices}
Starting with the set of Boltzmann weights (\ref{Boltzmann_weights})
satisfying the Yang-Baxter equation we can define a family of transfer
matrices generating the commuting integrals for a system of $SU(n)_k$ anyons
in the vector representation.
Introducing local inhomogeneities $\{u_k\}_{k=1}^L$ the first transfer matrix
for the $A_{n-1}^{(1)}$ IRF model is constructed as usual
\begin{align}
%\begin{equation}
\label{trnsf_mtx_period}
  \mathbf{T}(u) 
  & = \,\prod_{k = 1}^L 
  \Wop[a_{k-1}][a_k][b_{k-1}][b_{k}][u-u_k] 
  \nonumber\\
  &=
\begin{picture}(300,40)(-10,10)
%\put(-20,0){$=$}
\footnotesize
\put(0,0){\line(1,0){300}}
\put(0,0){\line(0,1){25}}
\put(0,25){\line(1,0){300}}
\put(300,0){\line(0,1){25}}
\put(50,0){\line(0,1){25}}
\put(250,0){\line(0,1){25}}
\put(125,0){\line(0,1){25}}
\put(175,0){\line(0,1){25}}
\put(13,10){$u-u_1$}
\put(83,10){$\cdots$}
\put(138,10){$u-u_k$}
\put(207,10){$\cdots$}
\put(263,10){$u-u_L$}
\put(-2,-10){$b_0$}
\put(47,-10){$b_1$}
\put(297,-10){$b_L$}
\put(247,-10){$b_{L-1}$}
\put(122,-10){$b_{k-1}$}
\put(172,-10){$b_k$}
\put(-2,30){$a_0$}
\put(47,30){$a_1$}
\put(122,30){$a_{k-1}$}
\put(172,30){$a_k$}
\put(247,30){$a_{L-1}$}
\put(297,30){$a_L$}
\put(25,0){\vector(-1,0){1}}
\put(85,0){\vector(-1,0){1}}
\put(150,0){\vector(-1,0){1}}
\put(210,0){\vector(-1,0){1}}
\put(275,0){\vector(-1,0){1}}
\put(25,25){\vector(-1,0){1}}
\put(85,25){\vector(-1,0){1}}
\put(150,25){\vector(-1,0){1}}
\put(210,25){\vector(-1,0){1}}
\put(275,25){\vector(-1,0){1}}
\put(0,12.5){\vector(0,1){1}}
\put(50,12.5){\vector(0,1){1}}
\put(125,12.5){\vector(0,1){1}}
\put(175,12.5){\vector(0,1){1}}
\put(250,12.5){\vector(0,1){1}}
\put(300,12.5){\vector(0,1){1}}
\put(0,-50){}
\end{picture}\\
\phantom{ }\nonumber
%\end{aligned}
%\end{equation}
\end{align}
after identifying $(a_L,b_L) = (a_0,b_0)$ for periodic boundary conditions.
The quantum Hamiltonian for the lattice model of $SU(n)_k$ anyons $\psi_{[1]}$
with local interactions is obtained from the homogeneous limit $u_k\to0$ for
all $k$ of the transfer matrix and can be expressed in terms of the
Yang-Baxter operators (\ref{YBop})
\begin{equation}
  H = \left.\partial_u \ln \mathbf{T}(u)\right|_{u=0} 
    = \sum_{i=1}^L \left.\partial_u \ln W_i(u)\right|_{u=0} \,.
\end{equation}

%%%%%%%%%%%%%%%%%%%%%%%%%%%%%%%%%%%%%%%%%%%%%%%%%%%%%%%%%%%%%%%%%%%%%%
To obtain the complete set of commuting integrals one has to consider fused
$A_{n-1}^{(1)}$ IRF models, similar as in the case of the integrable $SU(n)$
vertex models \cite{KuRe82,KuRe83}, see Appendix~\ref{app:vv} for $n=3$.
Analogous to (\ref{vv-utrans}) we define the second transfer matrix from this
family of operators acting on the space of fusion path states of $SU(n)_k$
anyons $\psi_{[1]}$ in the horizontal direction by
\begin{align}
  \mathbf{U}(u)
  & = P_1^- \, \mathbf{T}(u) \, \mathbf{T}(u - 1) \nonumber \\
  & = \sum_{b_i} 
  \Wop[a_0][b_{0}][b_{L}][c_0][-1]
  \prod_{k=1}^L
  \Wop[a_{k-1}][a_k][b_{k-1}][b_k][u-u_k] 
  \Wop[b_{k-1}][b_k][c_{k-1}][c_k][u-u_k-1]
  \label{irf-utrans} \\
  & =
\begin{picture}(300,50)(-60,-2)
\footnotesize
\put(0,0){\line(-1,1){25}}
\put(0,0){\line(-1,-1){25}}
\put(-25,-25){\line(-1,1){25}}
\put(-50,0){\line(1,1){25}}
\put(-32,-3){$-1$}
\put(0,0){\line(1,0){300}}
\put(0,-25){\line(1,0){300}}
\put(0,-25){\line(0,1){50}}
\put(0,25){\line(1,0){300}}
\put(300,-25){\line(0,1){50}}
\put(50,-25){\line(0,1){50}}
\put(250,-25){\line(0,1){50}}
\put(125,-25){\line(0,1){50}}
\put(175,-25){\line(0,1){50}}
\put(13,10){$u-u_1$}
\put(83,10){$\cdots$}
\put(138,10){$u-u_k$}
\put(207,10){$\cdots$}
\put(263,10){$u-u_L$}
\put(3,-15){$u-u_1-1$}
\put(83,-15){$\cdots$}
\put(128,-15){$u-u_k-1$}
\put(207,-15){$\cdots$}
\put(252,-15){$u-u_L-1$}
\put(-2,-35){$c_0$}
\put(47,-35){$c_1$}
\put(297,-35){$c_L$}
\put(247,-35){$c_{L-1}$}
\put(122,-35){$c_{k-1}$}
\put(172,-35){$c_k$}
\put(-2,30){$a_0$}
\put(47,30){$a_1$}
\put(122,30){$a_{k-1}$}
\put(172,30){$a_k$}
\put(247,30){$a_{L-1}$}
\put(297,30){$a_L$}
\put(25,-25){\vector(-1,0){1}}
\put(85,-25){\vector(-1,0){1}}
\put(150,-25){\vector(-1,0){1}}
\put(210,-25){\vector(-1,0){1}}
\put(275,-25){\vector(-1,0){1}}
\put(25,0){\vector(-1,0){1}}
\put(85,0){\vector(-1,0){1}}
\put(150,0){\vector(-1,0){1}}
\put(210,0){\vector(-1,0){1}}
\put(275,0){\vector(-1,0){1}}
\put(25,25){\vector(-1,0){1}}
\put(85,25){\vector(-1,0){1}}
\put(150,25){\vector(-1,0){1}}
\put(210,25){\vector(-1,0){1}}
\put(275,25){\vector(-1,0){1}}
\put(0,12.5){\vector(0,1){1}}
\put(50,12.5){\vector(0,1){1}}
\put(125,12.5){\vector(0,1){1}}
\put(175,12.5){\vector(0,1){1}}
\put(250,12.5){\vector(0,1){1}}
\put(300,12.5){\vector(0,1){1}}
\put(0,-12.5){\vector(0,1){1}}
\put(50,-12.5){\vector(0,1){1}}
\put(125,-12.5){\vector(0,1){1}}
\put(175,-12.5){\vector(0,1){1}}
\put(250,-12.5){\vector(0,1){1}}
\put(300,-12.5){\vector(0,1){1}}
\put(53,4){$b_1$}
\put(178,4){$b_k$}
\put(305,-1){$b_L$}
\put(-58,5){$b_L$}
\multiput(-25,25)(4,0){7}{\line(0,1){1}}
\multiput(-25,-25)(4,0){7}{\line(0,1){1}}
\put(-50,0){\circle*{4}}
\put(0,0){\circle*{4}}
\put(50,0){\circle*{4}}
\put(125,0){\circle*{4}}
\put(175,0){\circle*{4}}
\put(250,0){\circle*{4}}
\put(300,0){\circle*{4}}
\put(-12,12){\vector(-1,1){1}}
\put(-37,13){\vector(1,1){1}}
\put(-37,-13){\vector(-1,1){1}}
\put(-12,-12){\vector(1,1){1}}
\end{picture}
\nonumber \\[10pt]
\nonumber
\end{align}
where, again, $(a_L,c_L)=(a_0,c_0)$ for periodic boundary conditions.  Further
transfer matrices are written as
\begin{equation}
  \mathbf{T}_\ell(u) = P_{\ell-1}^- \mathbf{T}(u)\mathbf{T}(u-1)\ldots
                              \mathbf{T}(u-\ell+1)\,,\qquad
\end{equation}
for $\ell=3,\ldots,n$.
Note that $P_{n-1}^-$ projects on a one-dimensional space, therefore
$\mathbf{T}_{n}(u) \equiv \Delta(u)\,\mathbb{I}$, similar as the quantum
determinant (\ref{vv-qdet}) of the vertex model.  The function $\Delta(u)$
depends on the type of $SU(n)_k$ anyons considered, for the present case of
$\psi_{[1]}$ we find
\begin{equation}
  \label{irf-qdet}
  \Delta(u) = \prod_{k=1}^L \left([u-u_k+1]\,
              \prod_{\ell=1}^{n-1} [u-u_k-\ell]\right)\,.
              % [u-u_k-2]\,\cdots\, [u-u_k-(n-1)]\,.
\end{equation}

\subsection{Fusion of weights}
Alternatively, the fused $A_{n-1}^{(1)}$ IRF models can be constructed by
means of the fusion procedure \cite{JKMO88}.  Boltzmann weights
$W^{[\lambda][\mu]}$(u) with admissible pairs corresponding to fusion with the
anyon $\psi_{[\lambda]}$ ($\psi_{[\mu]}$) along horizontal (vertical) links
are obtained from the partition function for a suitably chosen rectangular
block of the elementary ones (\ref{Boltzmann_weights}) with properly shifted
spectral parameters by projection onto the Young diagram $[\lambda]$
($[\mu]$).
In the present context where the admissible fusion path states in the
horizontal direction are those of the $SU(n)_k$ anyons $\psi_{[1]}$ we can
restrict ourselves to the vertically fused Boltzmann weights
$W^{[1][\mu]}(u)$.  Specifically, we consider the case of
$[\mu]=[1^2]\equiv[1,1,0,\ldots]$ which can be obtained by fusion of two
elementary weights.

Following the prescription used in Ref.~\cite{AJMP96} the fused Boltzmann
weights are constructed graphically as follows (fused edges are represented by
double lines on the corresponding link, as before arrows indicate the
direction in which the constraint for admissible pairs is to be read)
\begin{align*}
  W^{[1][1^2]}
  \left(\left.\begin{array}{@{}cc@{}}
        a_2&b_2\\a_0&b_0\end{array}\right|\,u
  \right)
&\equiv \Wvop[a_2][b_2][a_0][b_0][u]
\\[24pt]
& 
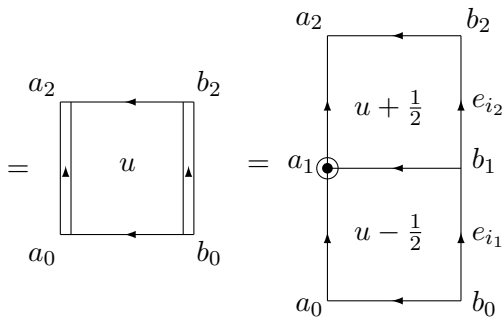
\begin{picture}(160,110)(-23,-40)
%
%\put(-110,22){$\Wvop[a_2][b_2][a_0][b_0][u] ~ = $}
\put(-20,22){$ = $}
\put(0,0){\line(1,0){50}}
\put(0,50){\line(1,0){50}}
\put(0,0){\line(0,1){50}}
\put(50,0){\line(0,1){50}}
\put(4,0){\line(0,1){50}}
\put(46,0){\line(0,1){50}}
\put(2,25){\vector(0,1){1}}
\put(48,25){\vector(0,1){1}}
\put(25,0){\vector(-1,0){1}}
\put(25,50){\vector(-1,0){1}}
\put(-12,54){$a_2$}
\put(51,54){$b_2$}
\put(-12,-10){$a_0$}
\put(51,-10){$b_0$}
\put(22,24){$u$}
\put(70,24){$=$}
\put(100,-25){\line(1,0){50}}
\put(100,25){\line(1,0){50}}
\put(100,75){\line(1,0){50}}
\put(100,-25){\line(0,1){100}}
\put(150,-25){\line(0,1){100}}
\put(100,25){\circle*{4}}
\put(100,25){\circle{8}}
\put(100,0){\vector(0,1){1}}
\put(100,50){\vector(0,1){1}}
\put(150,0){\vector(0,1){1}}
\put(150,50){\vector(0,1){1}}
\put(125,-25){\vector(-1,0){1}}
\put(125,25){\vector(-1,0){1}}
\put(125,75){\vector(-1,0){1}}
\put(88,79){$a_2$}
\put(85,25){$a_1$}
\put(88,-29){$a_0$}
\put(152,78){$b_2$}
\put(154,25){$b_1$}
\put(154,-29){$b_0$}
\put(110,45){$u+\frac{1}{2}$}
\put(110,-2){$u-\frac{1}{2}$}
\put(154,-2){$e_{i_1}$}
\put(154,48){$e_{i_2}$}
\end{picture}
\end{align*}
Here one has to perform an antisymmetrization on the nodes marked by $\odot$,
in addition to summation over all possible values, corresponding to the
projection onto $[1^2]$.  
The fused weights do not depend on the spin $b_1$.  They can be written in the
compact form
\begin{equation}
\begin{split}
& \Wvop[a + e_{i_k} + e_\Lambda][a + e_\Lambda][a + e_{i_k}][a][u] 
= - r_2(u) [u - \tfrac{1}{2}] [u + \tfrac{1}{2}] \prod_{j=1}^2 
\frac{[a_{i_ji_k} - 1]}{[a_{i_ji_k}]} \cr 
& \Wvop[a + e_{i_k} + e_\Lambda][a + e_\Lambda][a + e_{i_m}][a][u]
 = - r_2(u) [u- \tfrac{1}{2}]  [u+\tfrac{1}{2} + a_{i_ki_m}] 
 \frac{[1]}{[a_{i_ki_m}]} \prod_{\substack{j=1 \\ j\neq m}}^2 
\frac{[a_{i_ji_m}- 1]}{[a_{i_ji_m}]} 
 \cr 
& \Wvop[a + e_\Lambda + e_{i_m}][a + e_\Lambda][a + e_{i_m}][a][u] = 
- r_2(u) [u-\tfrac{1}{2}] [u+\tfrac{3}{2}]\prod_{\substack{j=1 \\ j\neq m}}^{2} 
\frac{[a_{i_ji_m} - 1]}{[a_{i_ji_m}]} \, ,\cr
\end{split}
\label{fused_weights}
\end{equation}
where we have defined $e_\Lambda = e_{i_1} + e_{i_2}$ and require that
$m\in\{1,2\}$ and $i_k \notin \{i_1, i_2\}$.  The spectral parameter dependent
scalar factor $r_2(u)$ is fixed by our choice of normalization of the
elementary weights (\ref{Boltzmann_weights}), see \cite{AJMP96}.

The antisymmetrization of the vertically fused weights allows to write them as
a column of two weights, multiplied by the projector $P_1^-$,
Eq.~(\ref{hecke-proj}), excluding straight paths in the adjacency graph.  In
terms of a graphical representation, we have
\begin{equation*}
\begin{picture}(250,75)(30,-30)

\put(45,-25){\line(1,0){50}}
\put(45,25){\line(1,0){50}}
\put(45,-25){\line(0,1){50}}
\put(95,-25){\line(0,1){50}}
\put(49,-25){\line(0,1){50}}
\put(91,-25){\line(0,1){50}}
\put(47,0){\vector(0,1){1}}
\put(93,0){\vector(0,1){1}}
\put(70,-25){\vector(-1,0){1}}
\put(70,25){\vector(-1,0){1}}
\put(33,29){$a_2$}
\put(96,29){$b_2$}
\put(33,-35){$a_0$}
\put(96,-35){$b_0$}
\put(56,-3){$u-\frac{1}{2}$}

\put(115,-2){$=$}

\put(150,0){\line(1,1){30}}
\put(150,0){\line(1,-1){30}}
\put(180,30){\line(1,-1){30}}
\put(180,-30){\line(1,1){30}}
\put(210,0){\circle*{4}}
\put(171,-2){$-1$}
\put(210,30){\line(0,-1){60}}
\put(210,30){\line(1,0){50}}
\put(210,0){\line(1,0){50}}
\put(210,-30){\line(1,0){50}}
\put(260,30){\line(0,-1){60}}
\multiput(180,30)(5,0){6}{\line(1,0){1}}
\multiput(180,-30)(5,0){6}{\line(1,0){1}}
\put(162,30){$a_2$}
%\put(136,-1){$b_1$}
\put(162,-30){$a_0$}
\put(214,5){$c$}
\put(266,30){$b_2$}
\put(266,-2){$b_1$}
\put(266,-32){$b_0$}
\put(235,10){$u$}
\put(223,-20){$u-1$}
\put(210,-15){\vector(0,1){1}}
\put(210,15){\vector(0,1){1}}
\put(260,-15){\vector(0,1){1}}
\put(260,15){\vector(0,1){1}}
\put(235,30){\vector(-1,0){1}}
\put(235,0){\vector(-1,0){1}}
\put(235,-30){\vector(-1,0){1}}
\put(195,-15){\vector(1,1){1}}
\put(165,-15){\vector(-1,1){1}}
\put(195,15){\vector(-1,1){1}}
\put(165,15){\vector(1,1){1}}

\end{picture}
\end{equation*}
%\vskip1.5cm
%
%\noindent 
This equivalence will turn out to be particularly useful when we consider the
transfer matrices and their algebraic relations in the next section.

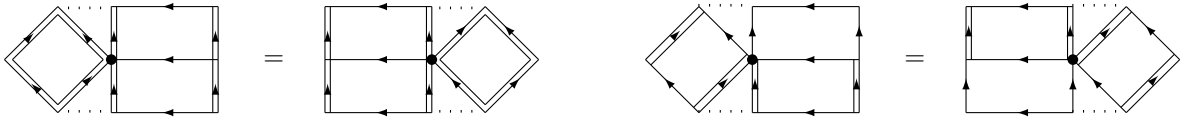
\begin{figure}[t]
\begin{picture}(400,90)(0,-10)
%%%%%%% Third YBE for vertically fused

% LHS
\put(0,0){\line(-1,1){20}}
\put(0,3){\line(-1,1){17}}
\put(0,0){\line(1,1){20}}
\put(0,3){\line(1,1){17}}
\put(-20,20){\line(1,1){20}}
\put(-17,20){\line(1,1){17}}
\put(0,40){\line(1,-1){20}}
\put(0,37){\line(1,-1){17}}

\put(20,20){\circle*{4}}
\put(20,20){\line(0,1){20}}
\put(22,20){\line(0,1){20}}
\put(20,20){\line(0,-1){20}}
\put(22,20){\line(0,-1){20}}
\put(20,40){\line(1,0){40}}
\put(20,20){\line(1,0){40}}
\put(20,0){\line(1,0){40}}
\put(60,20){\line(0,1){20}}
\put(58,20){\line(0,1){20}}
\put(60,20){\line(0,-1){20}}
\put(58,20){\line(0,-1){20}}

\multiput(4,0)(4,0){4}{\line(0,1){1}}
\multiput(4,39)(4,0){4}{\line(0,1){1}}

\put(40,0){\vector(-1,0){1}}
\put(40,20){\vector(-1,0){1}}
\put(40,40){\vector(-1,0){1}}
\put(59,10){\vector(0,1){1}}
\put(21,10){\vector(0,1){1}}
\put(59,30){\vector(0,1){1}}
\put(21,30){\vector(0,1){1}}

\put(12,13){\vector(1,1){1}}
\put(-9,10){\vector(-1,1){1}}
\put(10,29){\vector(-1,1){1}}
\put(-10,29){\vector(1,1){1}}

\put(77,18){$=$}

% RHS
\put(100,20){\line(0,1){20}}
\put(102,20){\line(0,1){20}}
\put(100,20){\line(0,-1){20}}
\put(102,20){\line(0,-1){20}}
\put(100,40){\line(1,0){40}}
\put(100,20){\line(1,0){40}}
\put(100,0){\line(1,0){40}}
\put(140,20){\line(0,1){20}}
\put(138,20){\line(0,1){20}}
\put(140,20){\line(0,-1){20}}
\put(138,20){\line(0,-1){20}}

\put(160,0){\line(-1,1){20}}
\put(160,3){\line(-1,1){17}}
\put(160,0){\line(1,1){20}}
\put(160,3){\line(1,1){17}}
\put(140,20){\line(1,1){20}}
\put(143,20){\line(1,1){17}}
\put(160,40){\line(1,-1){20}}
\put(160,37){\line(1,-1){17}}
\put(140,20){\circle*{4}}

\multiput(143,0)(4,0){4}{\line(0,1){1}}
\multiput(143,39)(4,0){4}{\line(0,1){1}}

\put(120,0){\vector(-1,0){1}}
\put(120,20){\vector(-1,0){1}}
\put(120,40){\vector(-1,0){1}}
\put(139,10){\vector(0,1){1}}
\put(101,10){\vector(0,1){1}}
\put(139,30){\vector(0,1){1}}
\put(101,30){\vector(0,1){1}}

\put(172,12){\vector(1,1){1}}
\put(148,12){\vector(-1,1){1}}
\put(168,32){\vector(-1,1){1}}
\put(152,32){\vector(1,1){1}}

%%%%%%% R TU = UT R

% LHS
\put(240,0){\line(-1,1){20}}
\put(240,0){\line(1,1){20}}
\put(238,2){\line(1,1){20}}
\put(220,20){\line(1,1){20}}
\put(222,18){\line(1,1){20}}
\put(240,40){\line(1,-1){20}}

\put(260,20){\circle*{4}}
\put(260,20){\line(0,1){20}}
\put(260,20){\line(0,-1){20}}
\put(262,20){\line(0,-1){20}}
\put(260,40){\line(1,0){40}}
%\put(260,38){\line(1,0){40}}
%\put(260,21){\line(1,0){40}}
\put(260,20){\line(1,0){40}}
%\put(260,2){\line(1,0){40}}
\put(260,0){\line(1,0){40}}
\put(300,20){\line(0,1){20}}
\put(300,20){\line(0,-1){20}}
\put(298,20){\line(0,-1){20}}

\multiput(240,0)(4,0){5}{\line(0,1){1}}
\multiput(240,40)(4,0){5}{\line(0,1){1}}

\put(280,0){\vector(-1,0){1}}
\put(280,20){\vector(-1,0){1}}
\put(280,40){\vector(-1,0){1}}
\put(299,12){\vector(0,1){1}}
\put(261,12){\vector(0,1){1}}
\put(300,32){\vector(0,1){1}}
\put(260,32){\vector(0,1){1}}

\put(250,12){\vector(1,1){1}}
\put(228,12){\vector(-1,1){1}}
\put(250,30){\vector(-1,1){1}}
\put(232,30){\vector(1,1){1}}

\put(317,18){$=$}

% RHS
\put(340,20){\line(0,1){20}}
\put(342,20){\line(0,1){20}}
\put(340,20){\line(0,-1){20}}
\put(340,40){\line(1,0){40}}
\put(340,20){\line(1,0){40}}
\put(340,0){\line(1,0){40}}
\put(380,20){\line(0,1){20}}
\put(378,20){\line(0,1){20}}
\put(380,20){\line(0,-1){20}}

\put(400,0){\line(-1,1){20}}
\put(400,0){\line(1,1){20}}
\put(398,2){\line(1,1){20}}
\put(380,20){\line(1,1){20}}
\put(382,18){\line(1,1){20}}
\put(400,40){\line(1,-1){20}}
\put(380,20){\circle*{4}}

\multiput(380,0)(4,0){5}{\line(0,1){1}}
\multiput(380,40)(4,0){5}{\line(0,1){1}}

\put(360,0){\vector(-1,0){1}}
\put(360,20){\vector(-1,0){1}}
\put(360,40){\vector(-1,0){1}}
\put(380,12){\vector(0,1){1}}
\put(340,12){\vector(0,1){1}}
\put(379,32){\vector(0,1){1}}
\put(341,32){\vector(0,1){1}}

\put(411,13){\vector(1,1){1}}
\put(388,12){\vector(-1,1){1}}
\put(408,32){\vector(-1,1){1}}
\put(393,31){\vector(1,1){1}}

\end{picture}
\caption{\label{fig:allybe}The additional Yang-Baxter relations containing the
  fused Boltzmann weights (\ref{fused_weights}).  Double lines represent
  fusion along the particular direction.  Indices and spectral parameters are
  suppressed for the sake of clarity.}
\end{figure}

As a final note, it should be stressed out that these fused weights satisfy a
set of Yang-Baxter relations, ensuring the preservation of integrability.  In
Figure~\ref{fig:allybe} we represent the  Yang-Baxter
relations satisfied by the elementary and fused Boltzmann weights in addition
to the initial one (\ref{YBE}).

%%%%%%%%%%%%%%%%%%%%%%%%%%%%%%%%%%%%%%%%%%%%%%%%%%%%%%%%%%%%%%%%%%%%%%
In terms of the vertically fused Boltzmann weights the transfer matrix
(\ref{irf-utrans}) is written as
\begin{equation}
\label{irf-utrans2}
\begin{aligned}
  \mathbf{U}(u)
& =
%\begin{picture}(300,65)(-60,10)
\begin{picture}(300,35)(-10,10)
\footnotesize

%\put(-25,11){$=$}

\put(0,0){\line(1,0){300}}
\put(2,0){\line(0,1){25}}
\put(0,0){\line(0,1){25}}
\put(0,25){\line(1,0){300}}
\put(300,0){\line(0,1){25}}
\put(298,0){\line(0,1){25}}
\put(50,0){\line(0,1){25}}
\put(52,0){\line(0,1){25}}
\put(250,0){\line(0,1){25}}
\put(248,0){\line(0,1){25}}
\put(123,0){\line(0,1){25}}
\put(125,0){\line(0,1){25}}
\put(175,0){\line(0,1){25}}
\put(177,0){\line(0,1){25}}

\put(3,10){$u-u_1-\frac{1}{2}$}
\put(83,10){$\cdots$}
\put(127,10){$u-u_k-\frac{1}{2}$}
\put(207,10){$\cdots$}
\put(250,10){$u-u_L-\frac{1}{2}$}

\put(-2,-10){$c_0$}
\put(47,-10){$c_1$}
\put(297,-10){$c_L$}
\put(247,-10){$c_{L-1}$}
\put(122,-10){$c_{k-1}$}
\put(172,-10){$c_k$}
\put(-2,30){$a_0$}
\put(47,30){$a_1$}
\put(122,30){$a_{k-1}$}
\put(172,30){$a_k$}
\put(247,30){$a_{L-1}$}
\put(297,30){$a_L$}

\put(25,0){\vector(-1,0){1}}
\put(85,0){\vector(-1,0){1}}
\put(150,0){\vector(-1,0){1}}
\put(210,0){\vector(-1,0){1}}
\put(275,0){\vector(-1,0){1}}
\put(25,25){\vector(-1,0){1}}
\put(85,25){\vector(-1,0){1}}
\put(150,25){\vector(-1,0){1}}
\put(210,25){\vector(-1,0){1}}
\put(275,25){\vector(-1,0){1}}

\put(1,12.5){\vector(0,1){1}}
\put(51,12.5){\vector(0,1){1}}
\put(124,12.5){\vector(0,1){1}}
\put(176,12.5){\vector(0,1){1}}
\put(249,12.5){\vector(0,1){1}}
\put(299,12.5){\vector(0,1){1}}

\end{picture}
\\[20pt]
&= \,\prod_{k = 1}^L 
  \Wvop[a_{k-1}][a_k][c_{k-1}][c_{k}][u-u_k-\frac12] \,,
\end{aligned}
\end{equation}
relating it to the fused $A_{n-1}^{(1)}$ IRF model based on fusion paths of
$\psi_{[1]}$ and $\psi_{[1^2]}$ anyons in the horizontal and vertical
direction, respectively.  Similarly, the transfer matrices
$\mathbf{T}_\ell(u)$ are obtained from the IRF model with $\psi_{[1^\ell]}$
anyons on the vertical links.  
We do not consider transfer matrices with anyons other than those related to
these fundamental representations on the vertical links here.  Based on our
results below we conjecture that they are algebraic functions of the
$\mathbf{T}_\ell(u)$, similar as in the integrable $SU(n)$ vertex models
\cite{KuRe82,KuRe83}, thus do not provide additional information on the
system.

Note that as a consequence of (\ref{fused_weights}) we have $\mathbf{U}(u_k+1)
= 0$ for $k=1,\ldots,L$, resembling Eq.~(\ref{vv-uzer}) for the vertex model.
This allows to extract a spectral parameter dependent factor from the second
transfer matrix
\begin{equation}
  \label{irf-ufac}
  \mathbf{U}(u) = \left(\prod_{k=1}^L\left[u-u_k-1\right]\right)\,
  \widetilde{\mathbf{U}}(u)\,.
\end{equation}
%similar as in the $SU(3)$ vertex model \cite{KuRe82}.

%%%%%%%%%%%%%%%%%%%%%%%%%%%%%%%%%%%%%%%%%%%%%%%%%%%%%%%%%%%%%%%%%%%%%% 

\section{Inversion identities for the $A_2^{(1)}$ IRF model}
%The face/vertex correspondence}
%
Motivated by the results for the $SU(3)$ vertex model given in the Appendix,
our aim is to transfer them to the face model.  Initially, the $A_1^{(1)}$ or
RSOS models have been introduced by Baxter in order to solve the eight-vertex
model \cite{Baxt73a,Baxt73b,Baxt73c}.  The observed correspondence between
Boltzmann weights of faces in the RSOS models and vertices in the eight-vertex
model turns out to be quite generic and can be formulated based on the
identification of equivalent operators in face and vertex models, respectively
\cite{Finch13}.  For the $A_{n-1}^{(1)}$ IRF models the face/vertex
correspondence was stated in Ref.~\cite{JKMO88}.

As we have seen above many of the quantities arising from the integrable
structures in the $SU(n)$ vertex models can be defined similarly in the
(fused) IRF models.  Inspired by this observation and the identities
(\ref{vv-set_invid}) satisfied by the transfer matrices of the $SU(3)$ vertex
model we now show hat similar relations hold for the transfer matrices
$\mathbf{T}(u)$ and $\mathbf{U}(u)$ of the inhomogeneous $A_2^{(1)}$ IRF
models, namely:
\begin{equation}
  \label{invid_face}
  \begin{aligned}
    & \mathbf{T}(u_k) \, \mathbf{T}(u_k - 1) = \mathbf{U}(u_k) \, , \cr
    & \mathbf{T}(u_k) \, \mathbf{U}(u_k-1) = \Delta(u_k) \, , \cr
    & \mathbf{U}(u_k) \, \mathbf{U}(u_k-1) = \Delta(u_k) \, \mathbf{T}(u_k -1
    ) \, , 
  \end{aligned}
\end{equation}
for $k=1,\ldots,L$.  Not all of these identities are independent: as in the
vertex model each relation for given $k$ can be obtained from the other two.
Furthermore, due to the translational invariance of the model, one has e.g.\
\begin{equation}
  \prod_{k = 1}^L \mathbf{T}(u_k) = \Big(
  \prod_{k,\ell = 1}^L [u_k - u_\ell + 1] 
  \Big) \mathbb{I} \, .
\end{equation}

For the proof of these identities we make use of the properties of the
Boltzmann weights introduced above adapted to the rank $n-1=2$ case: First, we
note that for the case of the $A_2^{(1)}$ model the anyon $\psi_{[1^2]}$
corresponds to the adjoint of the vector representation.  Therefore the
antisymmetrization in the fusion of Boltzmann weights effectively reverses the
order of local states in the admissible pairs or, graphically,
\begin{align*}
\begin{picture}(110,20)(0,0)
  \put(0,-1.5){\line(1,0){40}}
  \put(20,0){\vector(1,0){2}}
  \put(0, 1.5){\line(1,0){40}}
  \put(50,-3){$=$}
  \put(65,0){\line(1,0){40}}
  \put(85,0){\vector(-1,0){2}}
\end{picture}
\end{align*}
This can be used to show that, upon proper normalization of the $W_v$, we have
in particular
\begin{align}
  \label{wv-invid}
  \begin{picture}(70,70)(100,0)
    \put(100,-45){\line(1,0){50}}
    \put(100,05){\line(1,0){50}}
    \put(100,55){\line(1,0){50}}
    \put(100,-45){\line(0,1){100}}
    \put(150,-45){\line(0,1){100}}
    \put(104,-45){\line(0,1){50}}
    \put(146,-45){\line(0,1){50}}
    \put(100,5){\circle*{4}}
    \put(102,-20){\vector(0,1){1}}
    \put(100,30){\vector(0,1){1}}
    \put(148,-20){\vector(0,1){1}}
    \put(150,30){\vector(0,1){1}}
    \put(125,-45){\vector(-1,0){1}}
    \put(125,5){\vector(-1,0){1}}
    \put(125,55){\vector(-1,0){1}}
    \put(92,57){$a$}
    \put(90,5){$g$}
    \put(92,-49){$a$}
    \put(152,57){$d$}
    \put(153,3){$c$}
    \put(154,-49){$b$}
    \put(122,25){$u$}
    \put(110,-22){$u-\frac{3}{2}$}
  % \put(173,2){$= ~ \delta_{bd} \, 
  % \mathcal{C}_{ab} \, [u-2][u-1][u+1]\, ,$}
  %
  \end{picture}
  =\delta_{bd} \, \mathcal{C}_{ab} \,[u+1][u-1] [u-2]\, ,
  \\[20pt]
  \nonumber
\end{align}
where the coefficients $\mathcal{C}$ are given by 
\begin{equation}
\mathcal{C}_{ab} = 
\begin{cases}
 \frac{[1]}{[2]}\,  & \textrm{if}~~a~~ \textrm{is a corner state in the
   adjacency graph} \\
 \frac{[2]}{[1]}\,  & \textrm{if}~~b~~ \textrm{is a corner state in the
   adjacency graph} \\
 1\,  & \textrm{otherwise} \, .
\end{cases}
\end{equation}

Furthermore, since $\sum_{k=1}^n e_n=0$, the definition (\ref{fused_weights})
of the fused weights implies that, for $n=3$, they satisfy the initial
condition for $u=-\frac{3}{2}$
\begin{equation} 
  \Wvop[c][d][b][a][-\frac{3}{2}]  \sim \delta_{ac} \, .
\end{equation}
Note that a similar relation holds for the Boltzmann weights obtained by
fusing $n-1$ faces in the $SU(n)_k$ anyon model.
% and additional local relations, in particular unitarity and Yang-Baxter
% relations. 
% As will become transparent in the next section, unitarity plays a crucial
% role in proving the inversion identities.
%

\subsection{Proof of the inversion identities}
\label{sec:proof}

The first of the inversion identities (\ref{invid_face}) follows trivially
from the definition (\ref{irf-utrans}) and the fact that the $k$-th column of
$\mathbf{T}(u_k)\mathbf{T}(u_k-1)$ reduces to the projection operator
$P_1^-$.  Using the Yang-Baxter equation (\ref{YBE}) and $(P_1^-)^2=P_1^-$
yields the desired result.

To prove the second of the inversion identities (\ref{invid_face}), it is
useful to employ the expression of $\mathbf{U}$ in terms of the vertically
fused weights (\ref{irf-utrans2}) and exploit the inversion relation
(\ref{wv-invid}) satisfied by them.  Define also $u_{ij} = u_i-u_j$. We have
then
\begin{align*}
    \mathbf{T}(u_k) \, & \mathbf{U}(u_k-1) =\\
    & = 
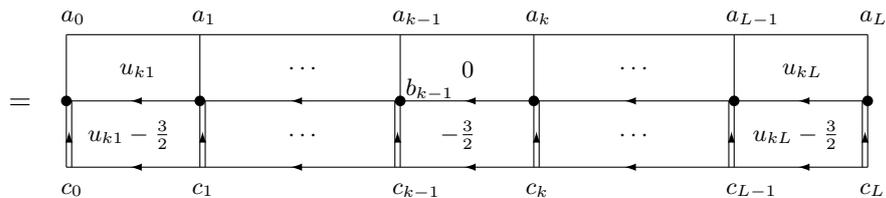
\begin{picture}(300,50)(-10,22)
% \vskip1.5cm
\footnotesize
%\put(-25,23){$=$}
%
\put(0,0){\line(1,0){300}}
\put(2,0){\line(0,1){25}}
\put(0,0){\line(0,1){50}}
\put(0,25){\line(1,0){300}}
\put(0,50){\line(1,0){300}}
\put(300,0){\line(0,1){50}}
\put(298,0){\line(0,1){25}}
\put(50,0){\line(0,1){50}}
\put(52,0){\line(0,1){25}}
\put(250,0){\line(0,1){50}}
\put(248,0){\line(0,1){25}}
\put(123,0){\line(0,1){25}}
\put(125,0){\line(0,1){50}}
\put(175,0){\line(0,1){50}}
\put(177,0){\line(0,1){25}}
\put(0,25){\circle*{4}}
\put(50,25){\circle*{4}}
\put(125,25){\circle*{4}}
\put(175,25){\circle*{4}}
\put(250,25){\circle*{4}}
\put(300,25){\circle*{4}}
\put(20,35){$u_{k1}$}
\put(83,35){$\cdots$}
\put(148,34){$0$}
\put(207,35){$\cdots$}
\put(268,35){$u_{kL}$}
\put(8,10){$u_{k1}-\frac{3}{2}$}
\put(83,10){$\cdots$}
\put(140,10){$-\frac{3}{2}$}
\put(207,10){$\cdots$}
\put(257,10){$u_{kL}-\frac{3}{2}$}
\put(-2,-10){$c_0$}
\put(47,-10){$c_1$}
\put(297,-10){$c_L$}
\put(247,-10){$c_{L-1}$}
\put(122,-10){$c_{k-1}$}
\put(172,-10){$c_k$}
\put(-2,55){$a_0$}
\put(47,55){$a_1$}
\put(122,55){$a_{k-1}$}
\put(172,55){$a_k$}
\put(247,55){$a_{L-1}$}
\put(297,55){$a_L$}
\put(127,27){$b_{k-1}$}
\put(25,0){\vector(-1,0){1}}
\put(85,0){\vector(-1,0){1}}
\put(150,0){\vector(-1,0){1}}
\put(210,0){\vector(-1,0){1}}
\put(275,0){\vector(-1,0){1}}
\put(25,25){\vector(-1,0){1}}
\put(85,25){\vector(-1,0){1}}
\put(150,25){\vector(-1,0){1}}
\put(210,25){\vector(-1,0){1}}
\put(275,25){\vector(-1,0){1}}
\put(1,12.5){\vector(0,1){1}}
\put(51,12.5){\vector(0,1){1}}
\put(124,12.5){\vector(0,1){1}}
\put(176,12.5){\vector(0,1){1}}
\put(249,12.5){\vector(0,1){1}}
\put(299,12.5){\vector(0,1){1}}
\end{picture}
\\[10pt]
\end{align*}
Directly from the definitions of the elementary Boltzmann weights
(\ref{Boltzmann_weights}) and the vertically fused ones (\ref{fused_weights}),
it follows that the above expression contains a factor
$\delta_{a_k,b_{k-1}}\,\delta_{c_k,b_{k-1}}\sim\delta_{a_kc_k}$.  Repeated use
of the inversion relation (\ref{wv-invid}) between $W_v$ and $W$ then leads to
\begin{equation}
  \begin{split}
    \mathbf{T}(u_k) \, \mathbf{U}(u_k-1) & = 
    \prod_{\ell=1}^L\,[ u_{k\ell}+1]\, [u_{k\ell}-1]\, [u_{k\ell}-2] \,
    \mathcal{C}_{c_\ell c_{\ell+1}}  \cr
    & = \Delta(u_k) \, \mathbb{I} \, ,
  \end{split}
\end{equation}
where we have used the fact that, as a consequence of periodic boundary
conditions for our model,
\begin{equation} 
\prod_{\ell=1}^L \mathcal{C}_{c_\ell c_{\ell+1}} = 1\,. %\mathbb{I} \, .
\end{equation}
Alternatively, the proof can be done using the projector operators discussed
in Section~\ref{sec:Hecke} (in fact, this form will turn out to be more
convenient in order to prove the third of the inversion identities below):
%
%\newpage
\begin{align*}
  \mathbf{T}(u_k) & \, \mathbf{U}(u_k-1) =\\
  &=
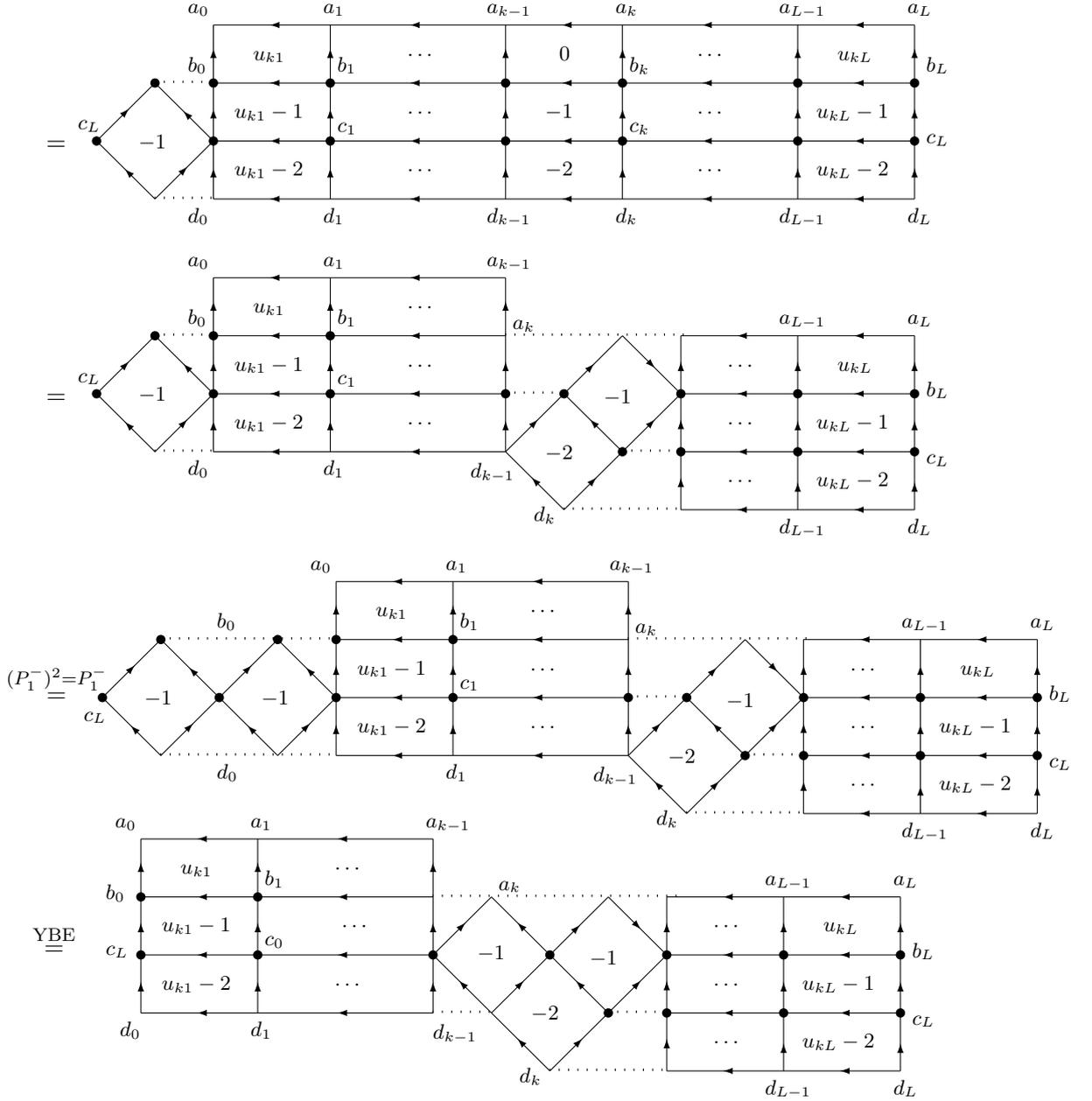
\begin{picture}(400,80)(-60,-4)
\footnotesize
\put(0,0){\line(-1,1){25}}
\put(0,0){\line(-1,-1){25}}
\put(-25,-25){\line(-1,1){25}}
\put(-50,0){\line(1,1){25}}
\put(-32,-3){$-1$}
\put(0,0){\line(1,0){300}}
\put(0,-25){\line(1,0){300}}
\put(0,-25){\line(0,1){75}}
\put(0,25){\line(1,0){300}}
\put(300,-25){\line(0,1){75}}
\put(50,-25){\line(0,1){75}}
\put(250,-25){\line(0,1){75}}
\put(125,-25){\line(0,1){75}}
\put(175,-25){\line(0,1){75}}
\put(0,50){\line(1,0){300}}
\put(17,35){$u_{k1}$}
\put(83,35){$\cdots$}
\put(148,35){$0$}
\put(207,35){$\cdots$}
\put(267,35){$u_{kL}$}
\put(9,10){$u_{k1}-1$}
\put(83,10){$\cdots$}
\put(142,10){$-1$}
\put(207,10){$\cdots$}
\put(258,10){$u_{kL}-1$}
\put(9,-15){$u_{k1}-2$}
\put(83,-15){$\cdots$}
\put(142,-15){$-2$}
\put(207,-15){$\cdots$}
\put(258,-15){$u_{kL}-2$}
\put(-11,55){$a_0$}
\put(47,55){$a_1$}
\put(297,55){$a_L$}
\put(242,55){$a_{L-1}$}
\put(117,55){$a_{k-1}$}
\put(172,55){$a_k$}
\put(-11,-35){$d_0$}
\put(47,-35){$d_1$}
\put(297,-35){$d_L$}
\put(242,-35){$d_{L-1}$}
\put(117,-35){$d_{k-1}$}
\put(172,-35){$d_k$}
\put(-11,30){$b_0$}
\put(53,30){$b_1$}
\put(178,30){$b_k$}
\put(305,30){$b_L$}
\put(25,-25){\vector(-1,0){1}}
\put(85,-25){\vector(-1,0){1}}
\put(150,-25){\vector(-1,0){1}}
\put(210,-25){\vector(-1,0){1}}
\put(275,-25){\vector(-1,0){1}}
\put(25,0){\vector(-1,0){1}}
\put(85,0){\vector(-1,0){1}}
\put(150,0){\vector(-1,0){1}}
\put(210,0){\vector(-1,0){1}}
\put(275,0){\vector(-1,0){1}}
\put(25,25){\vector(-1,0){1}}
\put(85,25){\vector(-1,0){1}}
\put(150,25){\vector(-1,0){1}}
\put(210,25){\vector(-1,0){1}}
\put(275,25){\vector(-1,0){1}}
\put(25,50){\vector(-1,0){1}}
\put(85,50){\vector(-1,0){1}}
\put(150,50){\vector(-1,0){1}}
\put(210,50){\vector(-1,0){1}}
\put(275,50){\vector(-1,0){1}}
\put(0,13){\vector(0,1){1}}
\put(50,13){\vector(0,1){1}}
\put(125,13){\vector(0,1){1}}
\put(175,13){\vector(0,1){1}}
\put(250,13){\vector(0,1){1}}
\put(300,13){\vector(0,1){1}}
\put(0,-12.5){\vector(0,1){1}}
\put(50,-12.5){\vector(0,1){1}}
\put(125,-12.5){\vector(0,1){1}}
\put(175,-12.5){\vector(0,1){1}}
\put(250,-12.5){\vector(0,1){1}}
\put(300,-12.5){\vector(0,1){1}}
\put(0,39){\vector(0,1){1}}
\put(50,39){\vector(0,1){1}}
\put(125,39){\vector(0,1){1}}
\put(175,39){\vector(0,1){1}}
\put(250,39){\vector(0,1){1}}
\put(300,39){\vector(0,1){1}}
\put(53,4){$c_1$}
\put(178,4){$c_k$}
\put(305,-1){$c_L$}
\put(-58,5){$c_L$}
\multiput(-25,25)(4,0){7}{\line(0,1){1}}
\multiput(-25,-25)(4,0){7}{\line(0,1){1}}
\put(-50,0){\circle*{4}}
\put(0,0){\circle*{4}}
\put(50,0){\circle*{4}}
\put(125,0){\circle*{4}}
\put(175,0){\circle*{4}}
\put(250,0){\circle*{4}}
\put(300,0){\circle*{4}}
\put(-25,25){\circle*{4}}
\put(0,25){\circle*{4}}
\put(50,25){\circle*{4}}
\put(125,25){\circle*{4}}
\put(175,25){\circle*{4}}
\put(250,25){\circle*{4}}
\put(300,25){\circle*{4}}
\put(-12,12){\vector(-1,1){1}}
\put(-37,13){\vector(1,1){1}}
\put(-37,-13){\vector(-1,1){1}}
\put(-12,-12){\vector(1,1){1}}
\end{picture}
\displaybreak[1]
\\
& =
\begin{picture}(400,100)(-60,-4)
\footnotesize
\put(0,0){\line(-1,1){25}}
\put(0,0){\line(-1,-1){25}}
\put(-25,-25){\line(-1,1){25}}
\put(-50,0){\line(1,1){25}}
\put(-32,-3){$-1$}
\put(0,0){\line(1,0){125}}
\put(0,-25){\line(1,0){125}}
\put(0,-25){\line(0,1){75}}
\put(0,25){\line(1,0){125}}
\put(300,-50){\line(0,1){75}}
\put(50,-25){\line(0,1){75}}
\put(250,-50){\line(0,1){75}}
\put(125,-25){\line(0,1){75}}
\put(200,-50){\line(0,1){75}}
\put(0,50){\line(1,0){125}}
\put(200,-50){\line(1,0){100}}
\put(200,0){\line(1,0){100}}
\put(200,25){\line(1,0){100}}
\put(200,-25){\line(1,0){100}}
%%%%%% face with -2
\put(125,-25){\line(1,1){25}}
\put(125,-25){\line(1,-1){25}}
\put(175,-25){\line(-1,1){25}}
\put(175,-25){\line(-1,-1){25}}
\put(163,-37){\vector(1,1){1}}
\put(137,-37){\vector(-1,1){1}}
\put(138,-12){\vector(1,1){1}}
\put(162,-12){\vector(-1,1){1}}
%%%%%%%
%%% face with -1
\put(175,-25){\line(1,1){25}}
\put(175,25){\line(1,-1){25}}
\put(175,25){\line(-1,-1){25}}
\put(162,12){\vector(1,1){1}}
\put(188,12){\vector(1,-1){1}}
\put(188,-12){\vector(1,1){1}}
%%%%%
%
\put(17,35){$u_{k1}$}
\put(83,35){$\cdots$}
\put(220,-40){$\cdots$}
\put(267,10){$u_{kL}$}
\put(9,10){$u_{k1}-1$}
\put(83,10){$\cdots$}
\put(167,-4){$-1$}
\put(220,10){$\cdots$}
\put(258,-15){$u_{kL}-1$}
\put(9,-15){$u_{k1}-2$}
\put(83,-15){$\cdots$}
\put(142,-28){$-2$}
\put(220,-15){$\cdots$}
\put(258,-40){$u_{kL}-2$}
\put(-11,55){$a_0$}
\put(47,55){$a_1$}
\put(297,30){$a_L$}
\put(242,30){$a_{L-1}$}
\put(117,55){$a_{k-1}$}
\put(128,28){$a_k$}
\put(-11,-35){$d_0$}
\put(47,-35){$d_1$}
\put(297,-60){$d_L$}
\put(242,-60){$d_{L-1}$}
\put(110,-36){$d_{k-1}$}
\put(137,-55){$d_k$}
\put(-11,30){$b_0$}
\put(53,30){$b_1$}
\put(305,-1){$b_L$}
\put(25,-25){\vector(-1,0){1}}
\put(85,-25){\vector(-1,0){1}}
\put(225,-25){\vector(-1,0){1}}
\put(275,-25){\vector(-1,0){1}}
\put(25,0){\vector(-1,0){1}}
\put(85,0){\vector(-1,0){1}}
\put(225,0){\vector(-1,0){1}}
\put(275,0){\vector(-1,0){1}}
\put(25,25){\vector(-1,0){1}}
\put(85,25){\vector(-1,0){1}}
\put(225,25){\vector(-1,0){1}}
\put(275,25){\vector(-1,0){1}}
\put(25,50){\vector(-1,0){1}}
\put(85,50){\vector(-1,0){1}}
\put(225,-50){\vector(-1,0){1}}
\put(275,-50){\vector(-1,0){1}}
\put(0,13){\vector(0,1){1}}
\put(50,13){\vector(0,1){1}}
\put(125,13){\vector(0,1){1}}
\put(200,-12){\vector(0,1){1}}
\put(250,13){\vector(0,1){1}}
\put(300,13){\vector(0,1){1}}
\put(0,-12.5){\vector(0,1){1}}
\put(50,-12.5){\vector(0,1){1}}
\put(125,-12.5){\vector(0,1){1}}
\put(200,-37){\vector(0,1){1}}
\put(250,-12.5){\vector(0,1){1}}
\put(300,-12.5){\vector(0,1){1}}
\put(0,39){\vector(0,1){1}}
\put(50,39){\vector(0,1){1}}
\put(125,39){\vector(0,1){1}}
\put(200,13){\vector(0,1){1}}
\put(250,-39){\vector(0,1){1}}
\put(300,-39){\vector(0,1){1}}
\put(53,4){$c_1$}
\put(305,-30){$c_L$}
\put(-58,5){$c_L$}
\multiput(-25,25)(4,0){7}{\line(0,1){1}}
\multiput(-25,-25)(4,0){7}{\line(0,1){1}}
\put(-50,0){\circle*{4}}
\put(0,0){\circle*{4}}
\put(50,0){\circle*{4}}
\put(125,0){\circle*{4}}
\put(150,0){\circle*{4}}
\put(200,0){\circle*{4}}
\put(250,0){\circle*{4}}
\put(300,-25){\circle*{4}}
\put(-25,25){\circle*{4}}
\put(0,25){\circle*{4}}
\put(50,25){\circle*{4}}
\put(175,-25){\circle*{4}}
\put(200,-25){\circle*{4}}
\put(250,-25){\circle*{4}}
\put(300,0){\circle*{4}}
\put(-12,12){\vector(-1,1){1}}
\put(-37,13){\vector(1,1){1}}
\put(-37,-13){\vector(-1,1){1}}
\put(-12,-12){\vector(1,1){1}}
\multiput(125,25)(4,0){20}{\line(0,1){1}}
\multiput(125,0)(4,0){7}{\line(0,1){1}}
\multiput(175,-25)(4,0){7}{\line(0,1){1}}
\multiput(150,-50)(4,0){13}{\line(0,1){1}}
\end{picture}
\displaybreak[1]
\\
&\hspace{-6mm}\stackrel{(P_1^-)^2=P_1^-}{=}
\begin{picture}(450,120)(-95,-2)
%%%%%%%%%%%%%%%%%%%%% THIRD EQUALITY %%%%%%%%%%%%%%%%%%%%%%%%%%
\footnotesize
\put(-50,0){\line(-1,1){25}}
\put(-50,0){\line(-1,-1){25}}
\put(-75,-25){\line(-1,1){25}}
\put(-100,0){\line(1,1){25}}
\put(-82,-3){$-1$}
\put(0,0){\line(-1,1){25}}
\put(0,0){\line(-1,-1){25}}
\put(-25,-25){\line(-1,1){25}}
\put(-50,0){\line(1,1){25}}
\put(-32,-3){$-1$}
\put(0,0){\line(1,0){125}}
\put(0,-25){\line(1,0){125}}
\put(0,-25){\line(0,1){75}}
\put(0,25){\line(1,0){125}}
\put(300,-50){\line(0,1){75}}
\put(50,-25){\line(0,1){75}}
\put(250,-50){\line(0,1){75}}
\put(125,-25){\line(0,1){75}}
\put(200,-50){\line(0,1){75}}
\put(0,50){\line(1,0){125}}
\put(200,-50){\line(1,0){100}}
\put(200,0){\line(1,0){100}}
\put(200,25){\line(1,0){100}}
\put(200,-25){\line(1,0){100}}
%%%%%% face with -2
\put(125,-25){\line(1,1){25}}
\put(125,-25){\line(1,-1){25}}
\put(175,-25){\line(-1,1){25}}
\put(175,-25){\line(-1,-1){25}}
\put(163,-37){\vector(1,1){1}}
\put(137,-37){\vector(-1,1){1}}
\put(138,-12){\vector(1,1){1}}
\put(162,-12){\vector(-1,1){1}}
%%%%%%%
%
%%% face with -1
\put(175,-25){\line(1,1){25}}
\put(175,25){\line(1,-1){25}}
\put(175,25){\line(-1,-1){25}}
\put(162,12){\vector(1,1){1}}
\put(188,12){\vector(1,-1){1}}
\put(188,-12){\vector(1,1){1}}
%%%%%
%
\put(17,35){$u_{k1}$}
\put(83,35){$\cdots$}
\put(220,-40){$\cdots$}
\put(267,10){$u_{kL}$}
\put(9,10){$u_{k1}-1$}
\put(83,10){$\cdots$}
\put(167,-4){$-1$}
\put(220,10){$\cdots$}
\put(258,-15){$u_{kL}-1$}
\put(9,-15){$u_{k1}-2$}
\put(83,-15){$\cdots$}
\put(142,-28){$-2$}
\put(220,-15){$\cdots$}
\put(258,-40){$u_{kL}-2$}
\put(-11,55){$a_0$}
\put(47,55){$a_1$}
\put(297,30){$a_L$}
\put(242,30){$a_{L-1}$}
\put(117,55){$a_{k-1}$}
\put(128,28){$a_k$}
\put(-51,-35){$d_0$}
\put(47,-35){$d_1$}
\put(297,-60){$d_L$}
\put(242,-60){$d_{L-1}$}
\put(110,-36){$d_{k-1}$}
\put(137,-55){$d_k$}
\put(-51,30){$b_0$}
\put(53,30){$b_1$}
\put(305,-1){$b_L$}
\put(25,-25){\vector(-1,0){1}}
\put(85,-25){\vector(-1,0){1}}
\put(225,-25){\vector(-1,0){1}}
\put(275,-25){\vector(-1,0){1}}
\put(25,0){\vector(-1,0){1}}
\put(85,0){\vector(-1,0){1}}
\put(225,0){\vector(-1,0){1}}
\put(275,0){\vector(-1,0){1}}
\put(25,25){\vector(-1,0){1}}
\put(85,25){\vector(-1,0){1}}
\put(225,25){\vector(-1,0){1}}
\put(275,25){\vector(-1,0){1}}
\put(25,50){\vector(-1,0){1}}
\put(85,50){\vector(-1,0){1}}
\put(225,-50){\vector(-1,0){1}}
\put(275,-50){\vector(-1,0){1}}
\put(0,13){\vector(0,1){1}}
\put(50,13){\vector(0,1){1}}
\put(125,13){\vector(0,1){1}}
\put(200,-12){\vector(0,1){1}}
\put(250,13){\vector(0,1){1}}
\put(300,13){\vector(0,1){1}}
\put(0,-12.5){\vector(0,1){1}}
\put(50,-12.5){\vector(0,1){1}}
\put(125,-12.5){\vector(0,1){1}}
\put(200,-37){\vector(0,1){1}}
\put(250,-12.5){\vector(0,1){1}}
\put(300,-12.5){\vector(0,1){1}}
\put(0,39){\vector(0,1){1}}
\put(50,39){\vector(0,1){1}}
\put(125,39){\vector(0,1){1}}
\put(200,13){\vector(0,1){1}}
\put(250,-39){\vector(0,1){1}}
\put(300,-39){\vector(0,1){1}}
\put(53,4){$c_1$}
\put(305,-30){$c_L$}
\put(-108,-10){$c_L$}
\multiput(-75,25)(4,0){19}{\line(0,1){1}}
\multiput(-75,-25)(4,0){19}{\line(0,1){1}}
\put(-100,0){\circle*{4}}
\put(-50,0){\circle*{4}}
\put(-75,25){\circle*{4}}
\put(0,0){\circle*{4}}
\put(50,0){\circle*{4}}
\put(125,0){\circle*{4}}
\put(150,0){\circle*{4}}
\put(200,0){\circle*{4}}
\put(250,0){\circle*{4}}
\put(300,-25){\circle*{4}}
\put(-25,25){\circle*{4}}
\put(0,25){\circle*{4}}
\put(50,25){\circle*{4}}
\put(175,-25){\circle*{4}}
\put(200,-25){\circle*{4}}
\put(250,-25){\circle*{4}}
\put(300,0){\circle*{4}}
\put(-62,12){\vector(-1,1){1}}
\put(-87,13){\vector(1,1){1}}
\put(-87,-13){\vector(-1,1){1}}
\put(-62,-12){\vector(1,1){1}}
\put(-12,12){\vector(-1,1){1}}
\put(-37,13){\vector(1,1){1}}
\put(-37,-13){\vector(-1,1){1}}
\put(-12,-12){\vector(1,1){1}}
\multiput(125,25)(4,0){20}{\line(0,1){1}}
\multiput(125,0)(4,0){7}{\line(0,1){1}}
\multiput(175,-25)(4,0){7}{\line(0,1){1}}
\multiput(150,-50)(4,0){13}{\line(0,1){1}}
\end{picture}
\displaybreak[1]
\\
&
\hspace{-2mm}\stackrel{\textrm{YBE}}{=}
 \begin{picture}(400,100)(-50,0)
%%%%%%%%%%%%%%%%%%%%%%% LAST EQUALITY %%%%%%%%%%%%%%%%%%%%%%%%%%%%
\footnotesize
\put(-25,0){\line(1,0){125}}
\put(-25,-25){\line(1,0){125}}
\put(-25,-25){\line(0,1){75}}
\put(-25,25){\line(1,0){125}}
\put(25,-25){\line(0,1){75}}
\put(100,-25){\line(0,1){75}}
\put(-25,50){\line(1,0){125}}
\put(300,-50){\line(0,1){75}}
\put(250,-50){\line(0,1){75}}
\put(200,-50){\line(0,1){75}}
\put(200,-50){\line(1,0){100}}
\put(200,0){\line(1,0){100}}
\put(200,25){\line(1,0){100}}
\put(200,-25){\line(1,0){100}}
%
%%%% face came from YBE -1
\put(150,0){\line(-1,1){25}}
\put(125,25){\line(-1,-1){25}}
\put(100,0){\line(1,-1){25}}
\put(119,-2){$-1$}
\put(112,-12){\vector(-1,1){1}}
\put(113,13){\vector(1,1){1}}
\put(137,13){\vector(-1,1){1}}
%%%%%%
%
%%%%%% face with -2
\put(125,-25){\line(1,1){25}}
\put(125,-25){\line(1,-1){25}}
\put(175,-25){\line(-1,1){25}}
\put(175,-25){\line(-1,-1){25}}
\put(163,-37){\vector(1,1){1}}
\put(137,-37){\vector(-1,1){1}}
\put(138,-12){\vector(1,1){1}}
\put(162,-12){\vector(-1,1){1}}
%%%%%%%
%
%%% face with -1
\put(175,-25){\line(1,1){25}}
\put(175,25){\line(1,-1){25}}
\put(175,25){\line(-1,-1){25}}
\put(162,12){\vector(1,1){1}}
\put(188,12){\vector(1,-1){1}}
\put(188,-12){\vector(1,1){1}}
%%%%%
%
\put(-8,35){$u_{k1}$}
\put(58,35){$\cdots$}
\put(220,-40){$\cdots$}
\put(267,10){$u_{kL}$}
\put(-16,10){$u_{k1}-1$}
\put(61,10){$\cdots$}
\put(167,-4){$-1$}
\put(220,10){$\cdots$}
\put(258,-15){$u_{kL}-1$}
\put(-16,-15){$u_{k1}-2$}
\put(58,-15){$\cdots$}
\put(142,-28){$-2$}
\put(220,-15){$\cdots$}
\put(258,-40){$u_{kL}-2$}
\put(-36,55){$a_0$}
\put(22,55){$a_1$}
\put(297,30){$a_L$}
\put(242,30){$a_{L-1}$}
\put(97,55){$a_{k-1}$}
\put(128,28){$a_k$}
\put(-34,-35){$d_0$}
\put(22,-35){$d_1$}
\put(297,-60){$d_L$}
\put(242,-60){$d_{L-1}$}
\put(100,-36){$d_{k-1}$}
\put(137,-55){$d_k$}
\put(-40,24){$b_0$}
\put(28,30){$b_1$}
\put(305,-1){$b_L$}
\put(0,-25){\vector(-1,0){1}}
\put(60,-25){\vector(-1,0){1}}
\put(225,-25){\vector(-1,0){1}}
\put(275,-25){\vector(-1,0){1}}
\put(0,0){\vector(-1,0){1}}
\put(60,0){\vector(-1,0){1}}
\put(225,0){\vector(-1,0){1}}
\put(275,0){\vector(-1,0){1}}
\put(0,25){\vector(-1,0){1}}
\put(60,25){\vector(-1,0){1}}
\put(225,25){\vector(-1,0){1}}
\put(275,25){\vector(-1,0){1}}
\put(0,50){\vector(-1,0){1}}
\put(60,50){\vector(-1,0){1}}
\put(225,-50){\vector(-1,0){1}}
\put(275,-50){\vector(-1,0){1}}
\put(-25,13){\vector(0,1){1}}
\put(25,13){\vector(0,1){1}}
\put(100,13){\vector(0,1){1}}
\put(200,-12){\vector(0,1){1}}
\put(250,13){\vector(0,1){1}}
\put(300,13){\vector(0,1){1}}
\put(-25,-12.5){\vector(0,1){1}}
\put(25,-12.5){\vector(0,1){1}}
\put(100,-12.5){\vector(0,1){1}}
\put(200,-37){\vector(0,1){1}}
\put(250,-12.5){\vector(0,1){1}}
\put(300,-12.5){\vector(0,1){1}}
\put(-25,39){\vector(0,1){1}}
\put(25,39){\vector(0,1){1}}
\put(100,39){\vector(0,1){1}}
\put(200,13){\vector(0,1){1}}
\put(250,-39){\vector(0,1){1}}
\put(300,-39){\vector(0,1){1}}
\put(28,4){$c_0$}
\put(305,-30){$c_L$}
\put(-40,0){$c_L$}
\put(-25,0){\circle*{4}}
\put(25,0){\circle*{4}}
\put(100,0){\circle*{4}}
\put(150,0){\circle*{4}}
\put(200,0){\circle*{4}}
\put(250,0){\circle*{4}}
\put(300,-25){\circle*{4}}
\put(-25,25){\circle*{4}}
\put(25,25){\circle*{4}}
\put(175,-25){\circle*{4}}
\put(200,-25){\circle*{4}}
\put(250,-25){\circle*{4}}
\put(300,0){\circle*{4}}
\multiput(100,25)(4,0){27}{\line(0,1){1}}
\multiput(175,-25)(4,0){7}{\line(0,1){1}}
\multiput(150,-50)(4,0){13}{\line(0,1){1}}
\multiput(100,-25)(4,0){7}{\line(0,1){1}}
\end{picture}\\[25pt]
\end{align*}
In the last step we have repetitively used the Yang-Baxter relation 
in order to push the left projector faces, $P_1^-$, to the right, and
exploited once again the fact that $(P_1^-)^2=P_1^-$. 
One immediately recognizes then the formation of the second projector
operator, $P_2^-$, constructed in Section~\ref{sec:Hecke}.  As already noticed
in that section, the action of $P_2^-$ on a sequence $(a_0, \ldots, a_3)$ is
proportional to $\delta_{a_0a_3}$, the exact factor of proportionality being
the one derived above through the use of the vertically fused weights.

For the third of Eqs.~(\ref{invid_face}) we have in the same spirit
\begin{align*}
  \mathbf{U}(u_k) &\, \mathbf{U}(u_k-1) = \\ 
  & =
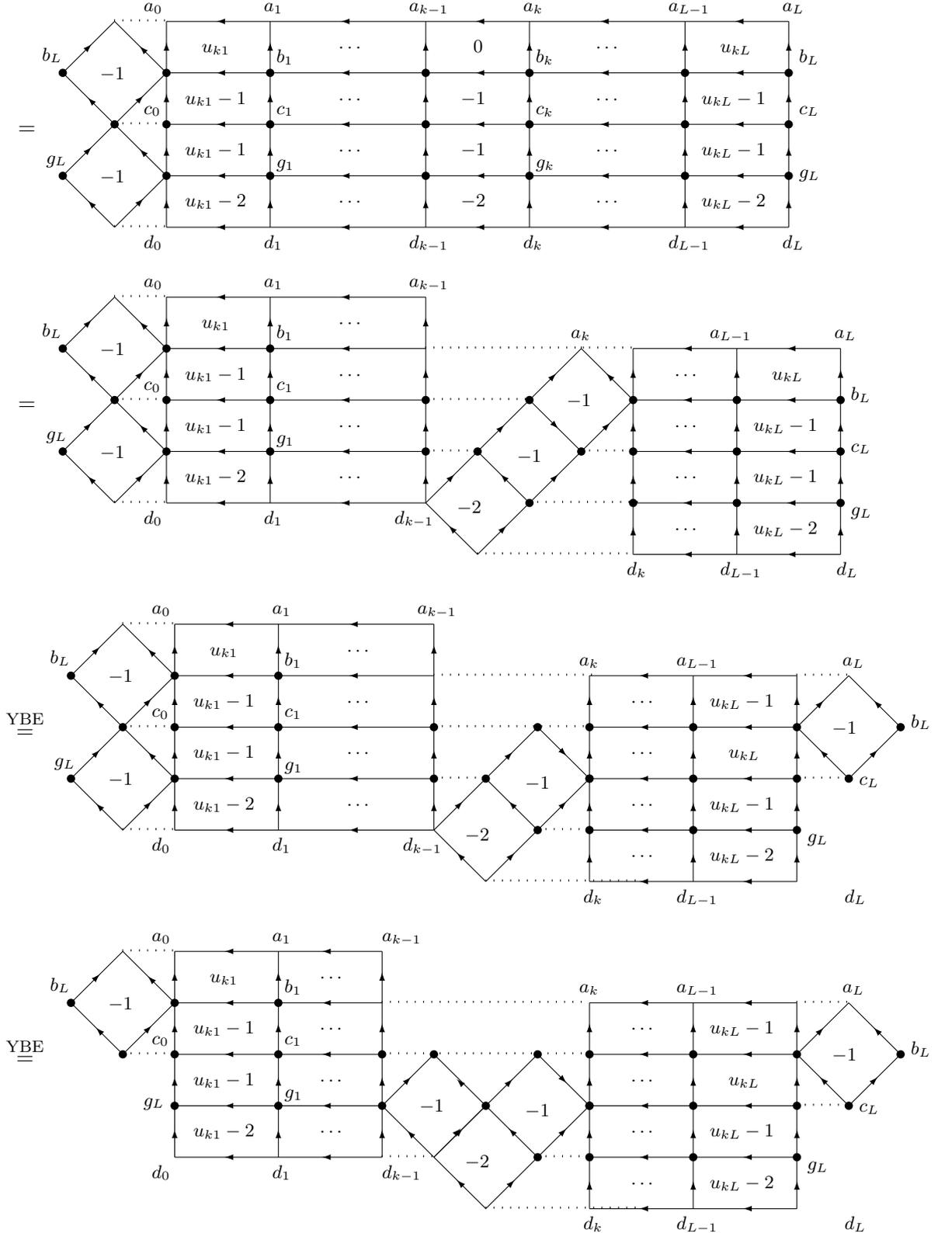
\begin{picture}(375,75)(-60,20)
\footnotesize
\put(0,0){\line(-1,1){25}}
\put(0,0){\line(-1,-1){25}}
\put(-25,-25){\line(-1,1){25}}
\put(-50,0){\line(1,1){25}}
\put(-32,-3){$-1$}
\put(-12,62){\vector(-1,1){1}}
\put(-37,63){\vector(1,1){1}}
\put(-37,37){\vector(-1,1){1}}
\put(-12,38){\vector(1,1){1}}
\put(0,50){\line(-1,1){25}}
\put(0,50){\line(-1,-1){25}}
\put(-25,25){\line(-1,1){25}}
\put(-50,50){\line(1,1){25}}
\put(-32,47){$-1$}
\put(-12,12){\vector(-1,1){1}}
\put(-37,13){\vector(1,1){1}}
\put(-37,-13){\vector(-1,1){1}}
\put(-12,-12){\vector(1,1){1}}
\put(0,0){\line(1,0){300}}
\put(0,-25){\line(1,0){300}}
\put(0,-25){\line(0,1){100}}
\put(0,25){\line(1,0){300}}
\put(300,-25){\line(0,1){100}}
\put(50,-25){\line(0,1){100}}
\put(250,-25){\line(0,1){100}}
\put(125,-25){\line(0,1){100}}
\put(175,-25){\line(0,1){100}}
\put(0,50){\line(1,0){300}}
\put(0,75){\line(1,0){300}}
\put(17,60){$u_{k1}$}
\put(83,60){$\cdots$}
\put(148,60){$0$}
\put(207,60){$\cdots$}
\put(267,60){$u_{kL}$}
\put(9,35){$u_{k1}-1$}
\put(83,35){$\cdots$}
\put(142,35){$-1$}
\put(207,35){$\cdots$}
\put(258,35){$u_{kL}-1$}
\put(9,10){$u_{k1}-1$}
\put(83,10){$\cdots$}
\put(142,10){$-1$}
\put(207,10){$\cdots$}
\put(258,10){$u_{kL}-1$}
\put(9,-15){$u_{k1}-2$}
\put(83,-15){$\cdots$}
\put(142,-15){$-2$}
\put(207,-15){$\cdots$}
\put(258,-15){$u_{kL}-2$}
\put(-11,80){$a_0$}
\put(47,80){$a_1$}
\put(297,80){$a_L$}
\put(242,80){$a_{L-1}$}
\put(117,80){$a_{k-1}$}
\put(172,80){$a_k$}
\put(-11,-35){$d_0$}
\put(47,-35){$d_1$}
\put(297,-35){$d_L$}
\put(242,-35){$d_{L-1}$}
\put(117,-35){$d_{k-1}$}
\put(172,-35){$d_k$}
\put(53,55){$b_1$}
\put(178,55){$b_k$}
\put(-60,56){$b_L$}
\put(305,55){$b_L$}
\put(-11,30){$c_0$}
\put(53,30){$c_1$}
\put(178,30){$c_k$}
\put(305,30){$c_L$}
\put(53,4){$g_1$}
\put(178,4){$g_k$}
\put(305,-1){$g_L$}
\put(-58,6){$g_L$}
\put(25,-25){\vector(-1,0){1}}
\put(25,0){\vector(-1,0){1}}
\put(25,25){\vector(-1,0){1}}
\put(25,50){\vector(-1,0){1}}
\put(25,75){\vector(-1,0){1}}
\put(85,-25){\vector(-1,0){1}}
\put(85,0){\vector(-1,0){1}}
\put(85,25){\vector(-1,0){1}}
\put(85,50){\vector(-1,0){1}}
\put(85,75){\vector(-1,0){1}}
\put(150,-25){\vector(-1,0){1}}
\put(150,0){\vector(-1,0){1}}
\put(150,25){\vector(-1,0){1}}
\put(150,50){\vector(-1,0){1}}
\put(150,75){\vector(-1,0){1}}
\put(210,0){\vector(-1,0){1}}
\put(210,-25){\vector(-1,0){1}}
\put(210,25){\vector(-1,0){1}}
\put(210,50){\vector(-1,0){1}}
\put(210,75){\vector(-1,0){1}}
\put(275,-25){\vector(-1,0){1}}
\put(275,0){\vector(-1,0){1}}
\put(275,25){\vector(-1,0){1}}
\put(275,50){\vector(-1,0){1}}
\put(275,75){\vector(-1,0){1}}
\put(0,64){\vector(0,1){1}}
\put(0,39){\vector(0,1){1}}
\put(0,13){\vector(0,1){1}}
\put(0,-12.5){\vector(0,1){1}}
\put(50,64){\vector(0,1){1}}
\put(50,39){\vector(0,1){1}}
\put(50,13){\vector(0,1){1}}
\put(50,-12.5){\vector(0,1){1}}
\put(125,64){\vector(0,1){1}}
\put(125,39){\vector(0,1){1}}
\put(125,13){\vector(0,1){1}}
\put(125,-12.5){\vector(0,1){1}}
\put(175,64){\vector(0,1){1}}
\put(175,39){\vector(0,1){1}}
\put(175,13){\vector(0,1){1}}
\put(175,-12.5){\vector(0,1){1}}
\put(250,64){\vector(0,1){1}}
\put(250,39){\vector(0,1){1}}
\put(250,13){\vector(0,1){1}}
\put(250,-12.5){\vector(0,1){1}}
\put(300,64){\vector(0,1){1}}
\put(300,39){\vector(0,1){1}}
\put(300,13){\vector(0,1){1}}
\put(300,-12.5){\vector(0,1){1}}
\multiput(-25,75)(4,0){7}{\line(0,1){1}}
\multiput(-25,25)(4,0){7}{\line(0,1){1}}
\multiput(-25,-25)(4,0){7}{\line(0,1){1}}
\put(-50,0){\circle*{4}}
\put(-50,50){\circle*{4}}
\put(-25,25){\circle*{4}}
\put(0,0){\circle*{4}}
\put(0,25){\circle*{4}}
\put(0,50){\circle*{4}}
\put(50,0){\circle*{4}}
\put(50,25){\circle*{4}}
\put(50,50){\circle*{4}}
\put(125,0){\circle*{4}}
\put(125,25){\circle*{4}}
\put(125,50){\circle*{4}}
\put(175,0){\circle*{4}}
\put(175,25){\circle*{4}}
\put(175,50){\circle*{4}}
\put(250,0){\circle*{4}}
\put(250,25){\circle*{4}}
\put(250,50){\circle*{4}}
\put(300,0){\circle*{4}}
\put(300,25){\circle*{4}}
\put(300,50){\circle*{4}}
\end{picture}
\displaybreak[1]
\\
&=
\begin{picture}(375,125)(-60,20)
%%%%%%%%%%%%%%%% 2nd EQUALITY %%%%%%%%%%%%%%%%%%%%%%%%%%%%%%%%%%
\footnotesize
\put(0,0){\line(-1,1){25}}
\put(0,0){\line(-1,-1){25}}
\put(-25,-25){\line(-1,1){25}}
\put(-50,0){\line(1,1){25}}
\put(-32,-3){$-1$}
\put(-12,62){\vector(-1,1){1}}
\put(-37,63){\vector(1,1){1}}
\put(-37,37){\vector(-1,1){1}}
\put(-12,38){\vector(1,1){1}}
\put(0,50){\line(-1,1){25}}
\put(0,50){\line(-1,-1){25}}
\put(-25,25){\line(-1,1){25}}
\put(-50,50){\line(1,1){25}}
\put(-32,47){$-1$}
\put(-12,12){\vector(-1,1){1}}
\put(-37,13){\vector(1,1){1}}
\put(-37,-13){\vector(-1,1){1}}
\put(-12,-12){\vector(1,1){1}}
\put(0,0){\line(1,0){125}}
\put(0,-25){\line(1,0){125}}
\put(0,-25){\line(0,1){100}}
\put(0,25){\line(1,0){125}}
\put(50,-25){\line(0,1){100}}
\put(125,-25){\line(0,1){100}}
\put(0,50){\line(1,0){125}}
\put(0,75){\line(1,0){125}}
\put(225,-50){\line(0,1){100}}
\put(275,-50){\line(0,1){100}}
\put(325,-50){\line(0,1){100}}
\put(225,-50){\line(1,0){100}}
\put(225,-25){\line(1,0){100}}
\put(225,0){\line(1,0){100}}
\put(225,25){\line(1,0){100}}
\put(225,50){\line(1,0){100}}
%
%%%%%% face with -2
\put(125,-25){\line(1,1){25}}
\put(125,-25){\line(1,-1){25}}
\put(175,-25){\line(-1,1){25}}
\put(175,-25){\line(-1,-1){25}}
\put(163,-37){\vector(1,1){1}}
\put(137,-37){\vector(-1,1){1}}
\put(138,-12){\vector(1,1){1}}
\put(162,-12){\vector(-1,1){1}}
\put(140,-30){$-2$}
%%%%%%%
%
%%% lower face with -1
\put(175,-25){\line(1,1){25}}
\put(175,25){\line(1,-1){25}}
\put(175,25){\line(-1,-1){25}}
\put(162,12){\vector(1,1){1}}
\put(188,12){\vector(1,-1){1}}
\put(188,-12){\vector(1,1){1}}
\put(169,-5){$-1$}
%%%%%
%
%%% upper face with -1
\put(225,25){\line(-1,-1){25}}
\put(175,25){\line(1,1){25}}
\put(200,50){\line(1,-1){25}}
\put(187,37){\vector(1,1){1}}
\put(213,37){\vector(-1,1){1}}
\put(213,13){\vector(1,1){1}}
\put(193,22){$-1$}
%%%%%
%
\multiput(125,0)(4,0){7}{\line(0,1){1}}
\multiput(125,25)(4,0){13}{\line(0,1){1}}
\multiput(125,50)(4,0){20}{\line(0,1){1}}
\multiput(150,-50)(4,0){19}{\line(0,1){1}}
\multiput(175,-25)(4,0){14}{\line(0,1){1}}
\multiput(200,0)(4,0){7}{\line(0,1){1}}
\multiput(200,50)(4,0){7}{\line(0,1){1}}
\put(17,60){$u_{k1}$}
\put(83,60){$\cdots$}
\put(245,35){$\cdots$}
\put(292,35){$u_{kL}$}
\put(9,35){$u_{k1}-1$}
\put(83,35){$\cdots$}
\put(245,10){$\cdots$}
\put(283,10){$u_{kL}-1$}
\put(9,10){$u_{k1}-1$}
\put(83,10){$\cdots$}
\put(245,-15){$\cdots$}
\put(283,-15){$u_{kL}-1$}
\put(9,-15){$u_{k1}-2$}
\put(83,-15){$\cdots$}
\put(245,-40){$\cdots$}
\put(283,-40){$u_{kL}-2$}
\put(-11,80){$a_0$}
\put(47,80){$a_1$}
\put(322,55){$a_L$}
\put(262,55){$a_{L-1}$}
\put(117,80){$a_{k-1}$}
\put(195,55){$a_k$}
\put(-11,-35){$d_0$}
\put(47,-35){$d_1$}
\put(323,-60){$d_L$}
\put(267,-60){$d_{L-1}$}
\put(110,-35){$d_{k-1}$}
\put(222,-60){$d_k$}
\put(53,55){$b_1$}
% \put(178,55){$b_k$}
\put(-60,56){$b_L$}
\put(330,25){$b_L$}
\put(-11,30){$c_0$}
\put(53,30){$c_1$}
% \put(178,30){$c_k$}
\put(330,0){$c_L$}
\put(53,4){$g_1$}
% \put(178,4){$g_k$}
\put(330,-31){$g_L$}
\put(-58,6){$g_L$}
\put(25,-25){\vector(-1,0){1}}
\put(25,0){\vector(-1,0){1}}
\put(25,25){\vector(-1,0){1}}
\put(25,50){\vector(-1,0){1}}
\put(25,75){\vector(-1,0){1}}
\put(85,-25){\vector(-1,0){1}}
\put(85,0){\vector(-1,0){1}}
\put(85,25){\vector(-1,0){1}}
\put(85,50){\vector(-1,0){1}}
\put(85,75){\vector(-1,0){1}}
\put(250,-50){\vector(-1,0){1}}
\put(250,-25){\vector(-1,0){1}}
\put(250,0){\vector(-1,0){1}}
\put(250,25){\vector(-1,0){1}}
\put(250,50){\vector(-1,0){1}}
\put(300,-50){\vector(-1,0){1}}
\put(300,-25){\vector(-1,0){1}}
\put(300,0){\vector(-1,0){1}}
\put(300,25){\vector(-1,0){1}}
\put(300,50){\vector(-1,0){1}}
\put(0,64){\vector(0,1){1}}
\put(0,39){\vector(0,1){1}}
\put(0,13){\vector(0,1){1}}
\put(0,-12.5){\vector(0,1){1}}
\put(50,64){\vector(0,1){1}}
\put(50,39){\vector(0,1){1}}
\put(50,13){\vector(0,1){1}}
\put(50,-12.5){\vector(0,1){1}}
\put(125,64){\vector(0,1){1}}
\put(125,39){\vector(0,1){1}}
\put(125,13){\vector(0,1){1}}
\put(125,-12.5){\vector(0,1){1}}
\put(225,39){\vector(0,1){1}}
\put(225,13){\vector(0,1){1}}
\put(225,-12.5){\vector(0,1){1}}
\put(225,-37.5){\vector(0,1){1}}
\put(275,39){\vector(0,1){1}}
\put(275,13){\vector(0,1){1}}
\put(275,-12.5){\vector(0,1){1}}
\put(275,-37.5){\vector(0,1){1}}
\put(325,39){\vector(0,1){1}}
\put(325,13){\vector(0,1){1}}
\put(325,-12.5){\vector(0,1){1}}
\put(325,-37.5){\vector(0,1){1}}
\multiput(-25,75)(4,0){7}{\line(0,1){1}}
\multiput(-25,25)(4,0){7}{\line(0,1){1}}
\multiput(-25,-25)(4,0){7}{\line(0,1){1}}
\put(-50,0){\circle*{4}}
\put(-50,50){\circle*{4}}
\put(-25,25){\circle*{4}}
\put(0,0){\circle*{4}}
\put(0,25){\circle*{4}}
\put(0,50){\circle*{4}}
\put(50,0){\circle*{4}}
\put(50,25){\circle*{4}}
\put(50,50){\circle*{4}}
\put(125,0){\circle*{4}}
\put(125,25){\circle*{4}}
\put(150,0){\circle*{4}}
\put(175,-25){\circle*{4}}
\put(175,25){\circle*{4}}
\put(200,0){\circle*{4}}
\put(225,-25){\circle*{4}}
\put(225,0){\circle*{4}}
\put(225,25){\circle*{4}}
\put(275,-25){\circle*{4}}
\put(275,0){\circle*{4}}
\put(275,25){\circle*{4}}
\put(325,-25){\circle*{4}}
\put(325,0){\circle*{4}}
\put(325,25){\circle*{4}}
\end{picture}
\displaybreak[1]
\\
&\hspace{-2mm}\stackrel{\textrm{YBE}}{=}
\begin{picture}(375,150)(-60,20)
%%%%%%%%%%%%%%%% 3rd EQUALITY %%%%%%%%%%%%%%%%%%%%%%%%%%%%%%%%%%
\footnotesize
\put(0,0){\line(-1,1){25}}
\put(0,0){\line(-1,-1){25}}
\put(-25,-25){\line(-1,1){25}}
\put(-50,0){\line(1,1){25}}
\put(-32,-3){$-1$}
\put(-12,62){\vector(-1,1){1}}
\put(-37,63){\vector(1,1){1}}
\put(-37,37){\vector(-1,1){1}}
\put(-12,38){\vector(1,1){1}}
\put(0,50){\line(-1,1){25}}
\put(0,50){\line(-1,-1){25}}
\put(-25,25){\line(-1,1){25}}
\put(-50,50){\line(1,1){25}}
\put(-32,47){$-1$}
\put(-12,12){\vector(-1,1){1}}
\put(-37,13){\vector(1,1){1}}
\put(-37,-13){\vector(-1,1){1}}
\put(-12,-12){\vector(1,1){1}}
\put(0,0){\line(1,0){125}}
\put(0,-25){\line(1,0){125}}
\put(0,-25){\line(0,1){100}}
\put(0,25){\line(1,0){125}}
\put(50,-25){\line(0,1){100}}
\put(125,-25){\line(0,1){100}}
\put(0,50){\line(1,0){125}}
\put(0,75){\line(1,0){125}}
\put(200,-50){\line(0,1){100}}
\put(250,-50){\line(0,1){100}}
\put(300,-50){\line(0,1){100}}
\put(200,-50){\line(1,0){100}}
\put(200,-25){\line(1,0){100}}
\put(200,0){\line(1,0){100}}
\put(200,25){\line(1,0){100}}
\put(200,50){\line(1,0){100}}
%
%%%%%% face with -2
\put(125,-25){\line(1,1){25}}
\put(125,-25){\line(1,-1){25}}
\put(175,-25){\line(-1,1){25}}
\put(175,-25){\line(-1,-1){25}}
\put(163,-37){\vector(1,1){1}}
\put(137,-37){\vector(-1,1){1}}
\put(138,-12){\vector(1,1){1}}
\put(162,-12){\vector(-1,1){1}}
\put(140,-30){$-2$}
%%%%%%%
%
%%% lower face with -1
\put(175,-25){\line(1,1){25}}
\put(175,25){\line(1,-1){25}}
\put(175,25){\line(-1,-1){25}}
\put(162,12){\vector(1,1){1}}
\put(188,12){\vector(1,-1){1}}
\put(188,-12){\vector(1,1){1}}
\put(169,-5){$-1$}
%%%%%
%
%%% upper face with -1 (moved right)
\put(350,25){\line(-1,-1){25}}
\put(300,25){\line(1,1){25}}
\put(300,25){\line(1,-1){25}}
\put(325,50){\line(1,-1){25}}
\put(312,37){\vector(1,1){1}}
\put(338,37){\vector(-1,1){1}}
\put(338,13){\vector(1,1){1}}
\put(312,13){\vector(-1,1){1}}
\put(317,22){$-1$}
\multiput(300,50)(4,0){7}{\line(0,1){1}}
\multiput(300,0)(4,0){7}{\line(0,1){1}}
%%%%%
%
\multiput(125,0)(4,0){7}{\line(0,1){1}}
\multiput(125,25)(4,0){19}{\line(0,1){1}}
\multiput(125,50)(4,0){20}{\line(0,1){1}}
\multiput(150,-50)(4,0){19}{\line(0,1){1}}
\multiput(175,-25)(4,0){7}{\line(0,1){1}}
\put(17,60){$u_{k1}$}
\put(83,60){$\cdots$}
\put(220,35){$\cdots$}
\put(258,35){$u_{kL}-1$}
\put(9,35){$u_{k1}-1$}
\put(83,35){$\cdots$}
\put(220,10){$\cdots$}
\put(267,10){$u_{kL}$}
\put(9,10){$u_{k1}-1$}
\put(83,10){$\cdots$}
\put(220,-15){$\cdots$}
\put(258,-15){$u_{kL}-1$}
\put(9,-15){$u_{k1}-2$}
\put(83,-15){$\cdots$}
\put(220,-40){$\cdots$}
\put(258,-40){$u_{kL}-2$}
\put(-11,80){$a_0$}
\put(47,80){$a_1$}
\put(322,55){$a_L$}
\put(242,55){$a_{L-1}$}
\put(117,80){$a_{k-1}$}
\put(195,55){$a_k$}
\put(-11,-35){$d_0$}
\put(47,-35){$d_1$}
\put(323,-60){$d_L$}
\put(242,-60){$d_{L-1}$}
\put(110,-35){$d_{k-1}$}
\put(197,-60){$d_k$}
\put(53,55){$b_1$}
\put(-60,56){$b_L$}
\put(355,25){$b_L$}
\put(-11,30){$c_0$}
\put(53,30){$c_1$}
% \put(178,30){$c_k$}
\put(330,-3){$c_L$}
\put(53,4){$g_1$}
% \put(178,4){$g_k$}
\put(305,-31){$g_L$}
\put(-58,6){$g_L$}
\put(25,-25){\vector(-1,0){1}}
\put(25,0){\vector(-1,0){1}}
\put(25,25){\vector(-1,0){1}}
\put(25,50){\vector(-1,0){1}}
\put(25,75){\vector(-1,0){1}}
\put(85,-25){\vector(-1,0){1}}
\put(85,0){\vector(-1,0){1}}
\put(85,25){\vector(-1,0){1}}
\put(85,50){\vector(-1,0){1}}
\put(85,75){\vector(-1,0){1}}
\put(225,-50){\vector(-1,0){1}}
\put(225,-25){\vector(-1,0){1}}
\put(225,0){\vector(-1,0){1}}
\put(225,25){\vector(-1,0){1}}
\put(225,50){\vector(-1,0){1}}
\put(275,-50){\vector(-1,0){1}}
\put(275,-25){\vector(-1,0){1}}
\put(275,0){\vector(-1,0){1}}
\put(275,25){\vector(-1,0){1}}
\put(275,50){\vector(-1,0){1}}
\put(0,64){\vector(0,1){1}}
\put(0,39){\vector(0,1){1}}
\put(0,13){\vector(0,1){1}}
\put(0,-12.5){\vector(0,1){1}}
\put(50,64){\vector(0,1){1}}
\put(50,39){\vector(0,1){1}}
\put(50,13){\vector(0,1){1}}
\put(50,-12.5){\vector(0,1){1}}
\put(125,64){\vector(0,1){1}}
\put(125,39){\vector(0,1){1}}
\put(125,13){\vector(0,1){1}}
\put(125,-12.5){\vector(0,1){1}}
\put(200,39){\vector(0,1){1}}
\put(200,13){\vector(0,1){1}}
\put(200,-12.5){\vector(0,1){1}}
\put(200,-37.5){\vector(0,1){1}}
\put(250,39){\vector(0,1){1}}
\put(250,13){\vector(0,1){1}}
\put(250,-12.5){\vector(0,1){1}}
\put(250,-37.5){\vector(0,1){1}}
\put(300,39){\vector(0,1){1}}
\put(300,13){\vector(0,1){1}}
\put(300,-12.5){\vector(0,1){1}}
\put(300,-37.5){\vector(0,1){1}}
\multiput(-25,75)(4,0){7}{\line(0,1){1}}
\multiput(-25,25)(4,0){7}{\line(0,1){1}}
\multiput(-25,-25)(4,0){7}{\line(0,1){1}}
\put(-50,0){\circle*{4}}
\put(-50,50){\circle*{4}}
\put(-25,25){\circle*{4}}
\put(0,0){\circle*{4}}
\put(0,25){\circle*{4}}
\put(0,50){\circle*{4}}
\put(50,0){\circle*{4}}
\put(50,25){\circle*{4}}
\put(50,50){\circle*{4}}
\put(125,0){\circle*{4}}
\put(125,25){\circle*{4}}
\put(150,0){\circle*{4}}
\put(175,-25){\circle*{4}}
\put(175,25){\circle*{4}}
\put(200,0){\circle*{4}}
\put(200,-25){\circle*{4}}
\put(200,0){\circle*{4}}
\put(200,25){\circle*{4}}
\put(250,-25){\circle*{4}}
\put(250,0){\circle*{4}}
\put(250,25){\circle*{4}}
\put(300,-25){\circle*{4}}
\put(300,0){\circle*{4}}
\put(300,25){\circle*{4}}
\put(325,0){\circle*{4}}
\put(350,25){\circle*{4}}
\end{picture}
\displaybreak[1]
\\
&\hspace{-2mm}\stackrel{\textrm{YBE}}{=}
\begin{picture}(375,150)(-60,20)
%%%%%%%%%%%%%%%% 4th EQUALITY %%%%%%%%%%%%%%%%%%%%%%%%%%%%%%%%%%
\footnotesize
%%%%% This is moved to the left to form the projector
\put(125,25){\line(-1,-1){25}}
\put(125,25){\line(1,-1){25}}
\put(125,-25){\line(-1,1){25}}
\put(125,-25){\line(1,1){25}}
\put(118,-3){$-1$}
\put(137,12){\vector(-1,1){1}}
\put(113,13){\vector(1,1){1}}
\put(113,-13){\vector(-1,1){1}}
\put(137,-12){\vector(1,1){1}}
%%%%%%
%
%%%%%% The left projector is still therre
\put(0,50){\line(-1,1){25}}
\put(0,50){\line(-1,-1){25}}
\put(-25,25){\line(-1,1){25}}
\put(-50,50){\line(1,1){25}}
\put(-32,47){$-1$}
\put(-12,62){\vector(-1,1){1}}
\put(-37,63){\vector(1,1){1}}
\put(-37,37){\vector(-1,1){1}}
\put(-12,38){\vector(1,1){1}}
%%%%%%%%
%
\put(0,0){\line(1,0){100}}
\put(0,-25){\line(1,0){100}}
\put(0,-25){\line(0,1){100}}
\put(0,25){\line(1,0){100}}
\put(50,-25){\line(0,1){100}}
\put(100,-25){\line(0,1){100}}
\put(0,50){\line(1,0){100}}
\put(0,75){\line(1,0){100}}
\put(200,-50){\line(0,1){100}}
\put(250,-50){\line(0,1){100}}
\put(300,-50){\line(0,1){100}}
\put(200,-50){\line(1,0){100}}
\put(200,-25){\line(1,0){100}}
\put(200,0){\line(1,0){100}}
\put(200,25){\line(1,0){100}}
\put(200,50){\line(1,0){100}}
%
%%%%%% face with -2
\put(125,-25){\line(1,1){25}}
\put(125,-25){\line(1,-1){25}}
\put(175,-25){\line(-1,1){25}}
\put(175,-25){\line(-1,-1){25}}
\put(163,-37){\vector(1,1){1}}
\put(137,-37){\vector(-1,1){1}}
\put(138,-12){\vector(1,1){1}}
\put(162,-12){\vector(-1,1){1}}
\put(140,-30){$-2$}
%%%%%%%
%
%%% lower face with -1
\put(175,-25){\line(1,1){25}}
\put(175,25){\line(1,-1){25}}
\put(175,25){\line(-1,-1){25}}
\put(162,12){\vector(1,1){1}}
\put(188,12){\vector(1,-1){1}}
\put(188,-12){\vector(1,1){1}}
\put(169,-5){$-1$}
%%%%%
%
%%% upper face with -1 (moved right)
\put(350,25){\line(-1,-1){25}}
\put(300,25){\line(1,1){25}}
\put(300,25){\line(1,-1){25}}
\put(325,50){\line(1,-1){25}}
\put(312,37){\vector(1,1){1}}
\put(338,37){\vector(-1,1){1}}
\put(338,13){\vector(1,1){1}}
\put(312,13){\vector(-1,1){1}}
\put(317,22){$-1$}
\multiput(300,50)(4,0){7}{\line(0,1){1}}
\multiput(300,0)(4,0){7}{\line(0,1){1}}
%%%%%
%
\multiput(100,-25)(4,0){7}{\line(0,1){1}}
\multiput(100,25)(4,0){25}{\line(0,1){1}}
\multiput(100,50)(4,0){25}{\line(0,1){1}}
\multiput(150,-50)(4,0){19}{\line(0,1){1}}
\multiput(175,-25)(4,0){7}{\line(0,1){1}}
\put(17,60){$u_{k1}$}
\put(70,60){$\cdots$}
\put(220,35){$\cdots$}
\put(258,35){$u_{kL}-1$}
\put(9,35){$u_{k1}-1$}
\put(70,35){$\cdots$}
\put(220,10){$\cdots$}
\put(267,10){$u_{kL}$}
\put(9,10){$u_{k1}-1$}
\put(70,10){$\cdots$}
\put(220,-15){$\cdots$}
\put(258,-15){$u_{kL}-1$}
\put(9,-15){$u_{k1}-2$}
\put(70,-15){$\cdots$}
\put(220,-40){$\cdots$}
\put(258,-40){$u_{kL}-2$}
\put(-11,80){$a_0$}
\put(47,80){$a_1$}
\put(322,55){$a_L$}
\put(242,55){$a_{L-1}$}
\put(100,80){$a_{k-1}$}
\put(195,55){$a_k$}
\put(-11,-35){$d_0$}
\put(47,-35){$d_1$}
\put(323,-60){$d_L$}
\put(242,-60){$d_{L-1}$}
\put(100,-35){$d_{k-1}$}
\put(197,-60){$d_k$}
\put(53,55){$b_1$}
\put(-60,56){$b_L$}
\put(355,25){$b_L$}
\put(-11,30){$c_0$}
\put(53,30){$c_1$}
% \put(178,30){$c_k$}
\put(330,-3){$c_L$}
\put(53,4){$g_1$}
\put(305,-31){$g_L$}
\put(-15,0){$g_L$}
\put(25,-25){\vector(-1,0){1}}
\put(25,0){\vector(-1,0){1}}
\put(25,25){\vector(-1,0){1}}
\put(25,50){\vector(-1,0){1}}
\put(25,75){\vector(-1,0){1}}
\put(72,-25){\vector(-1,0){1}}
\put(72,0){\vector(-1,0){1}}
\put(72,25){\vector(-1,0){1}}
\put(72,50){\vector(-1,0){1}}
\put(72,75){\vector(-1,0){1}}
\put(225,-50){\vector(-1,0){1}}
\put(225,-25){\vector(-1,0){1}}
\put(225,0){\vector(-1,0){1}}
\put(225,25){\vector(-1,0){1}}
\put(225,50){\vector(-1,0){1}}
\put(275,-50){\vector(-1,0){1}}
\put(275,-25){\vector(-1,0){1}}
\put(275,0){\vector(-1,0){1}}
\put(275,25){\vector(-1,0){1}}
\put(275,50){\vector(-1,0){1}}
\put(0,64){\vector(0,1){1}}
\put(0,39){\vector(0,1){1}}
\put(0,13){\vector(0,1){1}}
\put(0,-12.5){\vector(0,1){1}}
\put(50,64){\vector(0,1){1}}
\put(50,39){\vector(0,1){1}}
\put(50,13){\vector(0,1){1}}
\put(50,-12.5){\vector(0,1){1}}
\put(100,64){\vector(0,1){1}}
\put(100,39){\vector(0,1){1}}
\put(100,13){\vector(0,1){1}}
\put(100,-12.5){\vector(0,1){1}}
\put(200,39){\vector(0,1){1}}
\put(200,13){\vector(0,1){1}}
\put(200,-12.5){\vector(0,1){1}}
\put(200,-37.5){\vector(0,1){1}}
\put(250,39){\vector(0,1){1}}
\put(250,13){\vector(0,1){1}}
\put(250,-12.5){\vector(0,1){1}}
\put(250,-37.5){\vector(0,1){1}}
\put(300,39){\vector(0,1){1}}
\put(300,13){\vector(0,1){1}}
\put(300,-12.5){\vector(0,1){1}}
\put(300,-37.5){\vector(0,1){1}}
\multiput(-25,75)(4,0){7}{\line(0,1){1}}
\multiput(-25,25)(4,0){7}{\line(0,1){1}}
\put(-50,50){\circle*{4}}
\put(-25,25){\circle*{4}}
\put(0,0){\circle*{4}}
\put(0,25){\circle*{4}}
\put(0,50){\circle*{4}}
\put(50,0){\circle*{4}}
\put(50,25){\circle*{4}}
\put(50,50){\circle*{4}}
\put(100,0){\circle*{4}}
\put(100,25){\circle*{4}}
\put(125,25){\circle*{4}}
\put(150,0){\circle*{4}}
\put(175,-25){\circle*{4}}
\put(175,25){\circle*{4}}
\put(200,0){\circle*{4}}
\put(200,-25){\circle*{4}}
\put(200,0){\circle*{4}}
\put(200,25){\circle*{4}}
\put(250,-25){\circle*{4}}
\put(250,0){\circle*{4}}
\put(250,25){\circle*{4}}
\put(300,-25){\circle*{4}}
\put(300,0){\circle*{4}}
\put(300,25){\circle*{4}}
\put(325,0){\circle*{4}}
\put(350,25){\circle*{4}}
\end{picture}
\displaybreak[1]
\\
&\hspace{-2mm}\stackrel{\textrm{YBE}}{=}
\begin{picture}(375,150)(-60,20)
%%%%%%%%%%%%%%%% 5th EQUALITY %%%%%%%%%%%%%%%%%%%%%%%%%%%%%%%%%%
\footnotesize
%%%%% This is moved to the left to form the projector
\put(125,25){\line(-1,-1){25}}
\put(125,25){\line(1,-1){25}}
\put(125,-25){\line(-1,1){25}}
\put(125,-25){\line(1,1){25}}
\put(118,-3){$-1$}
\put(137,12){\vector(-1,1){1}}
\put(113,13){\vector(1,1){1}}
\put(113,-13){\vector(-1,1){1}}
\put(137,-12){\vector(1,1){1}}
%%%%%%
%
%%%%%% The left projector is still therre
\put(0,50){\line(-1,1){25}}
\put(0,50){\line(-1,-1){25}}
\put(-25,25){\line(-1,1){25}}
\put(-50,50){\line(1,1){25}}
\put(-32,47){$-1$}
\put(-12,62){\vector(-1,1){1}}
\put(-37,63){\vector(1,1){1}}
\put(-37,37){\vector(-1,1){1}}
\put(-12,38){\vector(1,1){1}}
%%%%%%%%
%
\put(0,0){\line(1,0){100}}
\put(0,-25){\line(1,0){100}}
\put(0,-25){\line(0,1){100}}
\put(0,25){\line(1,0){100}}
\put(50,-25){\line(0,1){100}}
\put(100,-25){\line(0,1){100}}
\put(0,50){\line(1,0){100}}
\put(0,75){\line(1,0){100}}
\put(200,-50){\line(0,1){100}}
\put(250,-50){\line(0,1){100}}
\put(300,-50){\line(0,1){100}}
\put(200,-50){\line(1,0){100}}
\put(200,-25){\line(1,0){100}}
\put(200,0){\line(1,0){100}}
\put(200,25){\line(1,0){100}}
\put(200,50){\line(1,0){100}}
%
%%%%%% face with -2
\put(125,-25){\line(1,1){25}}
\put(125,-25){\line(1,-1){25}}
\put(175,-25){\line(-1,1){25}}
\put(175,-25){\line(-1,-1){25}}
\put(163,-37){\vector(1,1){1}}
\put(137,-37){\vector(-1,1){1}}
\put(138,-12){\vector(1,1){1}}
\put(162,-12){\vector(-1,1){1}}
\put(140,-30){$-2$}
%%%%%%%
%
%%% lower face with -1
\put(175,-25){\line(1,1){25}}
\put(175,25){\line(1,-1){25}}
\put(175,25){\line(-1,-1){25}}
\put(162,12){\vector(1,1){1}}
\put(188,12){\vector(1,-1){1}}
\put(188,-12){\vector(1,1){1}}
\put(169,-5){$-1$}
%%%%%
%
\multiput(100,-25)(4,0){7}{\line(0,1){1}}
\multiput(100,25)(4,0){25}{\line(0,1){1}}
\multiput(100,50)(4,0){25}{\line(0,1){1}}
\multiput(150,-50)(4,0){19}{\line(0,1){1}}
\multiput(175,-25)(4,0){7}{\line(0,1){1}}
\put(17,60){$u_{k1}$}
\put(70,60){$\cdots$}
\put(220,35){$\cdots$}
\put(258,35){$u_{kL}-1$}
\put(9,35){$u_{k1}-1$}
\put(70,35){$\cdots$}
\put(220,10){$\cdots$}
\put(267,10){$u_{kL}$}
\put(9,10){$u_{k1}-1$}
\put(70,10){$\cdots$}
\put(220,-15){$\cdots$}
\put(258,-15){$u_{kL}-1$}
\put(9,-15){$u_{k1}-2$}
\put(70,-15){$\cdots$}
\put(220,-40){$\cdots$}
\put(258,-40){$u_{kL}-2$}
\put(-11,80){$a_0$}
\put(47,80){$a_1$}
\put(297,55){$a_L$}
\put(242,55){$a_{L-1}$}
\put(100,80){$a_{k-1}$}
\put(195,55){$a_k$}
\put(-11,-35){$d_0$}
\put(47,-35){$d_1$}
\put(297,-60){$d_L$}
\put(242,-60){$d_{L-1}$}
\put(100,-35){$d_{k-1}$}
\put(197,-60){$d_k$}
\put(53,55){$b_1$}
\put(-60,56){$b_L$}
\put(305,25){$b_L$}
\put(-11,30){$c_0$}
\put(53,30){$c_1$}
% \put(178,30){$c_k$}
\put(305,-3){$c_L$}
\put(53,4){$g_1$}
\put(305,-31){$g_L$}
\put(-15,0){$g_L$}
\put(25,-25){\vector(-1,0){1}}
\put(25,0){\vector(-1,0){1}}
\put(25,25){\vector(-1,0){1}}
\put(25,50){\vector(-1,0){1}}
\put(25,75){\vector(-1,0){1}}
\put(72,-25){\vector(-1,0){1}}
\put(72,0){\vector(-1,0){1}}
\put(72,25){\vector(-1,0){1}}
\put(72,50){\vector(-1,0){1}}
\put(72,75){\vector(-1,0){1}}
\put(225,-50){\vector(-1,0){1}}
\put(225,-25){\vector(-1,0){1}}
\put(225,0){\vector(-1,0){1}}
\put(225,25){\vector(-1,0){1}}
\put(225,50){\vector(-1,0){1}}
\put(275,-50){\vector(-1,0){1}}
\put(275,-25){\vector(-1,0){1}}
\put(275,0){\vector(-1,0){1}}
\put(275,25){\vector(-1,0){1}}
\put(275,50){\vector(-1,0){1}}
\put(0,64){\vector(0,1){1}}
\put(0,39){\vector(0,1){1}}
\put(0,13){\vector(0,1){1}}
\put(0,-12.5){\vector(0,1){1}}
\put(50,64){\vector(0,1){1}}
\put(50,39){\vector(0,1){1}}
\put(50,13){\vector(0,1){1}}
\put(50,-12.5){\vector(0,1){1}}
\put(100,64){\vector(0,1){1}}
\put(100,39){\vector(0,1){1}}
\put(100,13){\vector(0,1){1}}
\put(100,-12.5){\vector(0,1){1}}
\put(200,39){\vector(0,1){1}}
\put(200,13){\vector(0,1){1}}
\put(200,-12.5){\vector(0,1){1}}
\put(200,-37.5){\vector(0,1){1}}
\put(250,39){\vector(0,1){1}}
\put(250,13){\vector(0,1){1}}
\put(250,-12.5){\vector(0,1){1}}
\put(250,-37.5){\vector(0,1){1}}
\put(300,39){\vector(0,1){1}}
\put(300,13){\vector(0,1){1}}
\put(300,-12.5){\vector(0,1){1}}
\put(300,-37.5){\vector(0,1){1}}
\multiput(-25,75)(4,0){7}{\line(0,1){1}}
\multiput(-25,25)(4,0){7}{\line(0,1){1}}
\put(-50,50){\circle*{4}}
\put(-25,25){\circle*{4}}
\put(0,0){\circle*{4}}
\put(0,25){\circle*{4}}
\put(0,50){\circle*{4}}
\put(50,0){\circle*{4}}
\put(50,25){\circle*{4}}
\put(50,50){\circle*{4}}
\put(100,0){\circle*{4}}
\put(100,25){\circle*{4}}
\put(125,25){\circle*{4}}
\put(150,0){\circle*{4}}
\put(175,-25){\circle*{4}}
\put(175,25){\circle*{4}}
\put(200,0){\circle*{4}}
\put(200,-25){\circle*{4}}
\put(200,0){\circle*{4}}
\put(200,25){\circle*{4}}
\put(250,-25){\circle*{4}}
\put(250,0){\circle*{4}}
\put(250,25){\circle*{4}}
\put(300,-25){\circle*{4}}
\put(300,0){\circle*{4}}
\put(300,25){\circle*{4}}
\end{picture}
\displaybreak[1]
\\
&\hspace{-2mm}\stackrel{\textrm{YBE}}{=}
\begin{picture}(375,150)(-60,20)
%%%%%%%%%%%%%%%% 6th EQUALITY %%%%%%%%%%%%%%%%%%%%%%%%%%%%%%%%%%
\footnotesize
%
%\put(-40,23){$\stackrel{\textrm{YBE}}{=}$}
%%%%% This is moved to the left to form the projector
\put(125,25){\line(-1,-1){25}}
\put(125,25){\line(1,-1){25}}
\put(125,-25){\line(-1,1){25}}
\put(125,-25){\line(1,1){25}}
\put(118,-3){$-1$}
\put(137,12){\vector(-1,1){1}}
\put(113,13){\vector(1,1){1}}
\put(113,-13){\vector(-1,1){1}}
\put(137,-12){\vector(1,1){1}}
%%%%%%
%
%%%%%% The original upper projector moved to the right
\put(150,50){\line(-1,1){25}}
\put(150,50){\line(-1,-1){25}}
\put(125,25){\line(-1,1){25}}
\put(100,50){\line(1,1){25}}
\put(118,47){$-1$}
\put(138,62){\vector(-1,1){1}}
\put(113,63){\vector(1,1){1}}
\put(113,37){\vector(-1,1){1}}
\put(138,38){\vector(1,1){1}}
%%%%%%%%
%
\put(0,0){\line(1,0){100}}
\put(0,-25){\line(1,0){100}}
\put(0,-25){\line(0,1){100}}
\put(0,25){\line(1,0){100}}
\put(50,-25){\line(0,1){100}}
\put(100,-25){\line(0,1){100}}
\put(0,50){\line(1,0){100}}
\put(0,75){\line(1,0){100}}
\put(200,-50){\line(0,1){100}}
\put(250,-50){\line(0,1){100}}
\put(300,-50){\line(0,1){100}}
\put(200,-50){\line(1,0){100}}
\put(200,-25){\line(1,0){100}}
\put(200,0){\line(1,0){100}}
\put(200,25){\line(1,0){100}}
\put(200,50){\line(1,0){100}}
%
%%%%%% face with -2
\put(125,-25){\line(1,1){25}}
\put(125,-25){\line(1,-1){25}}
\put(175,-25){\line(-1,1){25}}
\put(175,-25){\line(-1,-1){25}}
\put(163,-37){\vector(1,1){1}}
\put(137,-37){\vector(-1,1){1}}
\put(138,-12){\vector(1,1){1}}
\put(162,-12){\vector(-1,1){1}}
\put(140,-30){$-2$}
%%%%%%%
%
%%% lower face with -1
\put(175,-25){\line(1,1){25}}
\put(175,25){\line(1,-1){25}}
\put(175,25){\line(-1,-1){25}}
\put(162,12){\vector(1,1){1}}
\put(188,12){\vector(1,-1){1}}
\put(188,-12){\vector(1,1){1}}
\put(169,-5){$-1$}
%%%%%
%
\multiput(100,-25)(4,0){7}{\line(0,1){1}}
\multiput(100,25)(4,0){25}{\line(0,1){1}}
\multiput(150,50)(4,0){13}{\line(0,1){1}}
\multiput(100,75)(4,0){6}{\line(0,1){1}}
\multiput(150,-50)(4,0){19}{\line(0,1){1}}
\multiput(175,-25)(4,0){7}{\line(0,1){1}}
\put(9,60){$u_{k1}-1$}
\put(70,60){$\cdots$}
\put(220,35){$\cdots$}
\put(258,35){$u_{kL}-1$}
\put(17,35){$u_{k1}$}
\put(70,35){$\cdots$}
\put(220,10){$\cdots$}
\put(267,10){$u_{kL}$}
\put(9,10){$u_{k1}-1$}
\put(70,10){$\cdots$}
\put(220,-15){$\cdots$}
\put(258,-15){$u_{kL}-1$}
\put(9,-15){$u_{k1}-2$}
\put(70,-15){$\cdots$}
\put(220,-40){$\cdots$}
\put(258,-40){$u_{kL}-2$}
\put(-11,80){$a_0$}
\put(47,80){$a_1$}
\put(297,55){$a_L$}
\put(242,55){$a_{L-1}$}
\put(100,80){$a_{k-1}$}
\put(195,55){$a_k$}
\put(-11,-35){$d_0$}
\put(47,-35){$d_1$}
\put(297,-60){$d_L$}
\put(242,-60){$d_{L-1}$}
\put(100,-35){$d_{k-1}$}
\put(197,-60){$d_k$}
\put(53,55){$b_1$}
\put(-14,49){$b_L$}
\put(305,25){$b_L$}
\put(-14,24){$c_0$}
\put(53,30){$c_1$}
% \put(178,30){$c_k$}
\put(305,-3){$c_L$}
\put(53,4){$g_1$}
\put(305,-31){$g_L$}
\put(-14,0){$g_L$}
\put(25,-25){\vector(-1,0){1}}
\put(25,0){\vector(-1,0){1}}
\put(25,25){\vector(-1,0){1}}
\put(25,50){\vector(-1,0){1}}
\put(25,75){\vector(-1,0){1}}
\put(72,-25){\vector(-1,0){1}}
\put(72,0){\vector(-1,0){1}}
\put(72,25){\vector(-1,0){1}}
\put(72,50){\vector(-1,0){1}}
\put(72,75){\vector(-1,0){1}}
\put(225,-50){\vector(-1,0){1}}
\put(225,-25){\vector(-1,0){1}}
\put(225,0){\vector(-1,0){1}}
\put(225,25){\vector(-1,0){1}}
\put(225,50){\vector(-1,0){1}}
\put(275,-50){\vector(-1,0){1}}
\put(275,-25){\vector(-1,0){1}}
\put(275,0){\vector(-1,0){1}}
\put(275,25){\vector(-1,0){1}}
\put(275,50){\vector(-1,0){1}}
\put(0,64){\vector(0,1){1}}
\put(0,39){\vector(0,1){1}}
\put(0,13){\vector(0,1){1}}
\put(0,-12.5){\vector(0,1){1}}
\put(50,64){\vector(0,1){1}}
\put(50,39){\vector(0,1){1}}
\put(50,13){\vector(0,1){1}}
\put(50,-12.5){\vector(0,1){1}}
\put(100,64){\vector(0,1){1}}
\put(100,39){\vector(0,1){1}}
\put(100,13){\vector(0,1){1}}
\put(100,-12.5){\vector(0,1){1}}
\put(200,39){\vector(0,1){1}}
\put(200,13){\vector(0,1){1}}
\put(200,-12.5){\vector(0,1){1}}
\put(200,-37.5){\vector(0,1){1}}
\put(250,39){\vector(0,1){1}}
\put(250,13){\vector(0,1){1}}
\put(250,-12.5){\vector(0,1){1}}
\put(250,-37.5){\vector(0,1){1}}
\put(300,39){\vector(0,1){1}}
\put(300,13){\vector(0,1){1}}
\put(300,-12.5){\vector(0,1){1}}
\put(300,-37.5){\vector(0,1){1}}
\put(0,0){\circle*{4}}
\put(0,25){\circle*{4}}
\put(0,50){\circle*{4}}
\put(50,0){\circle*{4}}
\put(50,25){\circle*{4}}
\put(50,50){\circle*{4}}
\put(100,0){\circle*{4}}
\put(100,25){\circle*{4}}
\put(100,50){\circle*{4}}
\put(125,25){\circle*{4}}
\put(150,0){\circle*{4}}
\put(175,-25){\circle*{4}}
\put(175,25){\circle*{4}}
\put(200,0){\circle*{4}}
\put(200,-25){\circle*{4}}
\put(200,0){\circle*{4}}
\put(200,25){\circle*{4}}
\put(250,-25){\circle*{4}}
\put(250,0){\circle*{4}}
\put(250,25){\circle*{4}}
\put(300,-25){\circle*{4}}
\put(300,0){\circle*{4}}
\put(300,25){\circle*{4}}
\end{picture}
\displaybreak[1]
\\
&\hspace{-4mm}\stackrel{\textrm{inv. id.}}{=} \Delta(u_k)
\begin{picture}(375,110)(-25,60)
%%%%%%%%%%%%%%%% 7th EQUALITY %%%%%%%%%%%%%%%%%%%%%%%%%%%%%%%%%%
\footnotesize
%\put(-60,58){$\stackrel{\textrm{inv. id.}}{=} \Delta(u_k)$}
%%%%%% The original upper projector moved to the right
\put(150,50){\line(-1,1){25}}
\put(150,50){\line(-1,-1){25}}
\put(125,25){\line(-1,1){25}}
\put(100,50){\line(1,1){25}}
\put(118,47){$-1$}
\put(138,62){\vector(-1,1){1}}
\put(113,63){\vector(1,1){1}}
\put(113,37){\vector(-1,1){1}}
\put(138,38){\vector(1,1){1}}
%%%%%%%%
%
\put(0,50){\line(0,1){25}}
\put(50,50){\line(0,1){25}}
\put(100,50){\line(0,1){25}}
\put(0,50){\line(1,0){100}}
\put(0,75){\line(1,0){100}}
\put(180,25){\line(0,1){25}}
\put(230,25){\line(0,1){25}}
\put(280,25){\line(0,1){25}}
\put(180,25){\line(1,0){100}}
\put(180,50){\line(1,0){100}}
\multiput(100,75)(4,0){6}{\line(0,1){1}}
\multiput(150,50)(4,0){9}{\line(0,1){1}}
\multiput(125,25)(4,0){15}{\line(0,1){1}}
\put(9,60){$u_{k1}-1$}
\put(70,60){$\cdots$}
\put(200,35){$\cdots$}
\put(238,35){$u_{kL}-1$}
\put(-9,80){$a_0$}
\put(47,80){$a_1$}
\put(277,55){$a_L$}
\put(222,55){$a_{L-1}$}
\put(100,80){$a_{k-1}$}
\put(175,55){$a_k$}
\put(-9,40){$d_0$}
\put(47,40){$d_1$}
\put(175,15){$d_k$}
\put(222,15){$d_{L-1}$}
\put(277,15){$d_L$}
\put(25,50){\vector(-1,0){1}}
\put(25,75){\vector(-1,0){1}}
\put(72,50){\vector(-1,0){1}}
\put(72,75){\vector(-1,0){1}}
\put(205,25){\vector(-1,0){1}}
\put(205,50){\vector(-1,0){1}}
\put(255,25){\vector(-1,0){1}}
\put(255,50){\vector(-1,0){1}}
\put(0,64){\vector(0,1){1}}
\put(50,64){\vector(0,1){1}}
\put(100,64){\vector(0,1){1}}
\put(180,39){\vector(0,1){1}}
\put(230,39){\vector(0,1){1}}
\put(280,39){\vector(0,1){1}}
\end{picture}
\displaybreak[1]
\\
&= \ \Delta(u_k)
\begin{picture}(375,80)(-30,60)
%%%%%%%%%%%%%%%% 8th EQUALITY %%%%%%%%%%%%%%%%%%%%%%%%%%%%%%%%%%
\footnotesize
%
%\put(-52,58){$= \ \Delta(u_k)$}
%
\put(0,50){\line(0,1){25}}
\put(50,50){\line(0,1){25}}
\put(100,50){\line(0,1){25}}
\put(0,50){\line(1,0){280}}
\put(0,75){\line(1,0){280}}
\put(180,50){\line(0,1){25}}
\put(230,50){\line(0,1){25}}
\put(280,50){\line(0,1){25}}
\put(9,60){$u_{k1}-1$}
\put(70,60){$\cdots$}
\put(115,60){$u_k - u_k - 1$}
\put(200,60){$\cdots$}
\put(238,60){$u_{kL}-1$}
\put(-11,80){$a_0$}
\put(47,80){$a_1$}
\put(277,80){$a_L$}
\put(222,80){$a_{L-1}$}
\put(100,80){$a_{k-1}$}
\put(175,80){$a_k$}
\put(-11,40){$d_0$}
\put(47,40){$d_1$}
\put(100,40){$d_{k-1}$}
\put(175,40){$d_k$}
\put(222,40){$d_{L-1}$}
\put(277,40){$d_L$}
\put(25,50){\vector(-1,0){1}}
\put(25,75){\vector(-1,0){1}}
\put(72,50){\vector(-1,0){1}}
\put(72,75){\vector(-1,0){1}}
\put(142,75){\vector(-1,0){1}}
\put(142,50){\vector(-1,0){1}}
\put(205,75){\vector(-1,0){1}}
\put(205,50){\vector(-1,0){1}}
\put(255,75){\vector(-1,0){1}}
\put(255,50){\vector(-1,0){1}}
\put(0,64){\vector(0,1){1}}
\put(50,64){\vector(0,1){1}}
\put(100,64){\vector(0,1){1}}
\put(180,62){\vector(0,1){1}}
\put(230,62){\vector(0,1){1}}
\put(280,62){\vector(0,1){1}}
\end{picture}
\\[25pt]
&= \ \Delta(u_k) \, \mathbf{T}(u_k - 1) \, .
\end{align*}
% 

%%%%%%%%%%%%%%%%%%%%%%%%%%%%%%%%%%%%%%%%%%%%%%%%%%%%%%%%%%%%%%%%%%%%%%
\section{Discussion}
In this paper we have proven a set of discrete inversion identities
(\ref{invid_face}) satisfied by transfer matrices of inhomogeneous versions of
an $SU(3)_k$ anyon chain (or $A_2^{(1)}$ IRF model).  As shown in the
appendix, related identities can be derived using the quantum inverse
scattering method (QISM) for the $SU(3)$ quantum spin chain (or vertex model).

Similar identities have been obtained for the six-vertex model within
Sklyanin's separation of variables solution and for the RSOS models starting
from local properties of the corresponding Boltzmann weights
\cite{Skly92,FrKa14}.  In these cases, both with underlying rank-$1$ quantum
group, the inversion identities, when complemented by information on the
analytical properties of the transfer matrix, have been found to provide a
formulation of the spectral problem which allows to compute all eigenvalues.
We emphasize, however, that such purely functional approach has to be
complemented by an independent check that a given solution actually
corresponds to an eigenvalue of the quantum chain.  For the six-vertex model
the latter is provided by the SoV approach.

To what extend similar results can be established for the rank $n-1=2$ models
considered here remains to be studied: for the $SU(3)$ vertex model some
evidence exists from the SoV approach \cite{Skly93} but various open
questions, e.g.\ concerning the actual construction of the separated variables
in the fundamental spin representation (which is necessary to compute
eigenstates) and the nature of their common spectrum, remain to be addressed.
In addition, eigenvalues obtained within the functional approach for the
vertex model can be checked, for the periodic boundary conditions considered
here, against those from the algebraic Bethe ansatz \cite{KuRe82}.

For the IRF model we have some preliminary numerical results for small
systems, but there remains the practical issue that identities such as
(\ref{invid_face}) (or (\ref{vv-set_invid}) for the vertex model) are not well
suited for an efficient computation of the transfer matrix eigenvalues.
Reversing the line of arguments used to obtain the inversion identities in the
appendix, however, they can be related to generalized TQ-equations such as
(\ref{vv-TQ}) for the $SU(3)$ vertex model.  This requires to find a
factorization of (\ref{irf-qdet}) compatible with the asymptotic behaviour or
the transfer matrices.
Part of the additional input required to address this question for the
critical IRF models is available: 
by definition the transfer matrices appearing in the inversion identities are
Fourier polynomials in the spectral parameter.  The underlying fusion algebra
allows to split the spectrum into topological sectors \cite{FrZu90} where the
asymptotics of the transfer matrix can given in terms of the eigenvalues of
the adjacency matrix (\ref{adjmat}).
As a consequence, the effect of the anyonic statistics on the spectrum of the
IRF model as compared to the vertex case is similar to that of a twist in the
boundary conditions (depending on the sector), in agreement with the results
for the $A_n^{(1)}$ IRF models obtained from the fusion
procedure \cite{BaRe90}.

Finally let us note that -- while we have concentrated in this paper on the
derivation of inversion identities for the the critical rank-$2$ IRF models
with generic inhomogeneities and subject to periodic boundary conditions -- we
expect that similar identities can be derived for the transfer matrices
$\mathbf{T}_{\ell}(u)$, $\ell=1,\ldots,n-1$, for the $A_{n-1}^{(1)}$ IRF model
and for models with open boundary conditions \cite{BFKZ96} -- similar as in
the case of the $SU(n)$ vertex models \cite{CYSW14a}.

We plan to
address some of these questions in future work.

%%%%%%%%%%%%%%%%%%%%%%%%%%%%%%%%%%%%%%%%%%%%%%%%%%%%%%%%%%%%%%%%%%%%%%
%%%%%%%%%%%%%%%%%%%%%%%%%%%%%%%%%%%%%%%%%%%%%%%%%%%%%%%%%%%%%%%%%%%%%%
%%%%%%%%%%%%%%%%%%%%%%%%%%%%%%%%%%%%%%%%%%%%%%%%%%%%%%%%%%%%%%%%%%%%%%
%%%%%%%%%%%%%%%%%%%%%%%%%%%%%%%%%%%%%%%%%%%%%%%%%%%%%%%%%%%%%%%%%%%%%%
\begin{acknowledgments}
This work has been supported by the Deutsche Forschungsgemeinschaft under
grant no. Fr~737/7.
\end{acknowledgments}

\appendix
\section{The $SU(3)$ vertex model}
\label{app:vv}
In this appendix we recall the structures underlying the integrability of the
$SU(3)$ invariant quantum spin chains and related two-dimensional vertex
models and their Bethe ansatz solutions
\cite{Suth75,KuRe82,KuRe83,Skly93}.
The Hilbert space of the vertex models is the tensor product
$\mathcal{H}=\otimes_{j=1}^L V_j$.  To be specific we consider the case where
$V_j\simeq \mathbb{C}^3$ is the space of quantum states at site $j$ of the
lattice corresponding to the fundamental three-dimensional (vector)
representation of $SU(3)$ (corresponding to the Young diagram $[1]=[1,0]$).
In the framework of the quantum inverse scattering method (QISM) we define the
monodromy matrix acting on the tensor product of the auxiliary space
$V_a\simeq\mathbb{C}^3$ and the Hilbert space $\mathcal{H}$ of the model as
\begin{equation}
  \label{vv-monod}
  \mathcal{T}_a(u) = \mathcal{L}_{aL}(u-u_L) \mathcal{L}_{a,L-1}(u-u_{L-1})
  \cdots \mathcal{L}_{a1}(u-u_1) \,.
\end{equation}
Here the $\mathcal{L}_{aj}(u)$ are operators acting non-trivially only on
$V_a\otimes V_j$.  They are given as
\begin{equation}
  \label{vv-lop}
  \mathcal{L}_{aj}(u) = u\, \mathbb{I} \otimes \mathbb{I} + 
      \sum_{k,\ell=1}^3 e_{k\ell}^{(a)}\otimes e_{\ell k}^{(j)}
\end{equation}
with $3\times3$ matrices $\left(e_{k\ell}^{(\alpha)}\right)_{mn} = \delta_{km}
\delta_{\ell n}$ acting on $V_\alpha$.  The complex parameters $u_j$,
$j=1,\ldots,L$, define inhomogeneities in the lattice.

The monodromy matrix (as well as the local $\mathcal{L}$-operators) is a
representation of the Yangian $\mathcal{Y}(su(3))$
\begin{equation}
  \label{vv-ybe}
  R_{ab}(u-v) \mathcal{T}_a(u) \mathcal{T}_b(v)
  = \mathcal{T}_b(v)  \mathcal{T}_a(u) R_{ab}(u-v)\,.
\end{equation}
The $R$-matrix comprises the structure constants of this algebra
\begin{equation} 
R_{ab}(u) = u \,\mathbb{I} \otimes \mathbb{I} +   \, \mathbb{P}_{ab} \, ,
\end{equation}
where $\mathbb{P}_{ab}$ is the permutation operator acting on the tensor
product $\mathbb{C}^3\otimes \mathbb{C}^3$ as $\mathbb{P}_{ab} (x\otimes y) =
y \otimes x\, , \forall x,y \in \mathbb{C}^3$.
As a consequence of the Yang-Baxter relations (\ref{vv-ybe}) the transfer
matrix
\begin{equation}
  \label{vv-trans}
  T(u) = \mathrm{tr}_a [\mathcal{T}_a(u)]\,
\end{equation}
forms a family of commuting operators, $ [T(u)\, , \, T(v) ] = 0$.  Starting
from the reference state $\otimes_{j=1}^L\left| 0 \right\rangle_j$, where each
spin is in the highest or lowest weight state of the local $SU(3)$ irrep, the
spectrum of the transfer matrices can be obtained by means of the nested
(coordinate or) algebraic Bethe ansatz (ABA) \cite{Yang67,Suth75,KuRe83}.

An alternative solution of the spectral problem relies on functional relations
between (\ref{vv-trans}) and more general $SU(3)$ symmetric transfer matrices
acting on the Hilbert space $\mathcal{H}$ of the vertex model
\cite{KuRS81,KuRe82}.
% generated by $\mathcal{L}$-operators in auxiliary spaces $V_a$ 
% corresponding to different representations of $SU(3)$:
%
With the projectors
\begin{equation} 
\label{vv-proj}
\begin{split}
 P_{ab}^- & = \frac{1}{2}(\mathbb{I} \otimes \mathbb{I} -\mathbb{P}_{ab}) 
 = -\frac12 R(-1) \\
 P_{abc}^- & = \frac16 (\mathbb{I} \otimes \mathbb{I} \otimes \mathbb{I}
 + \mathbb{P}_{ab}\mathbb{P}_{bc} + \mathbb{P}_{bc}\mathbb{P}_{ab} - 
 \mathbb{P}_{ab} - \mathbb{P}_{ac} - \mathbb{P}_{bc} )\, ,
\end{split}
\end{equation}
onto the antisymmetric subspaces of the product $V_a\otimes V_b$ and
$V_a\otimes V_b\otimes V_c$, $V_\alpha\simeq \mathbb{C}^3$, respectively, we
define a second transfer matrix $U(u)$ 
\begin{equation}
\label{vv-utrans}
  U(u) = \mathrm{tr}_{ab}\left[
    P_{ab}^-\, \mathcal{T}_a(u-1) \mathcal{T}_b(u) \right]
\end{equation}
generated by $\mathcal{L}$-operators corresponding to the adjoint $[1^2]$ of
the vector representation of $SU(3)$ in auxiliary space $V_a\simeq
\mathbb{C}^3$ and, similarly, the quantum determinant of the monodromy matrix
$\mathcal{T}(u)$,
\begin{equation} 
  \label{vv-qdet}
  \Delta(u) = \textrm{tr}_{abc} \big[ P_{abc}^- \, \mathcal{T}_a(u-2) \, 
  \mathcal{T}_b(u-1) \, \mathcal{T}_c(u) \big]\, 
\end{equation}
which generates the center of the Yangian $\mathcal{Y}(su(3))$.  By
construction $\Delta(u)$ is a polynomial in the spectral parameter.  It has
$c$-number valued coefficients and can be factorized as
\begin{equation}
  \Delta(u) = d_1(u-2) d_2(u-1) d_3(u)\, \mathbb{I}\,.
\end{equation}
The polynomials $d_j(u)$ depend on the representation of the Yangian in
question.  Here, i.e.\ for the inhomogeneous model (\ref{vv-monod}) based on
the vector representation of $SU(3)$ in all components of the quantum space,
they are found to be \cite{KuRe82}
\begin{equation}
\label{vv-ddd}
   d_1(u) = \prod_{j=1}^L (u-u_j) = d_2(u)\,,\quad
    d_3(u) = \prod_{j=1}^L (u-u_j+1)\,.
\end{equation}

The transfer matrices $T(u)$ and $U(u)$ generate the complete set of commuting
integrals of the $SU(3)$ spin chain.  They satisfy functional equations with
auxiliary operators $Q_{1,2}(u)$ \cite{KuRe82,Skly93,PrSt00}
\begin{equation}
\label{vv-TQ}
  \begin{aligned}
    & d_2(u-2)\,d_3(u-1)\, Q_1(u-3) - U(u-1) Q_1(u-2) \\
    & \qquad \qquad
    + d_1(u-2)\,T(u-1) Q_1(u-1) - d_1(u-1)\,d_1(u-2)\, Q_1(u) = 0 \,,
    \\ %-----------------------------------------------------------
    & d_3(u-2)\,d_3(u-1)\,Q_2(u-3) - d_3(u-1)\,T(u-2) Q_2(u-2) \\
    &  \qquad\qquad
    + U(u-1) Q_2(u-1)  - d_1(u-2)\,d_2(u-1)\, Q_2(u) =0\,,
  \end{aligned}
\end{equation}
% \begin{equation}
% \label{vv-TQi}
%   \begin{aligned}
%    % & u^L (u+1)^L Q_1(u-1) - (u+1)^L T(u-1) Q_1(u) \\
%    % & \qquad \qquad
%    % + U(u) Q_1(u+1) - (u-1)^L u^L Q_1(u+2) = 0 \,,
%     & u^L (u+1)^L Q_1(u-1) - (u+1)^L T(u-1) Q_1(u) \\
%     & \qquad \qquad
%     + U(u) Q_1(u+1) - (u-1)^L u^L Q_1(u+2) = 0 \,,
%     \\ %-----------------------------------------------------------
%     & d_1(u) d_2(u-1) Q_2(u-2) -U(u) Q_2(u-1) \\
%     &  \qquad\qquad
%     + d_3(u-1) T(u)Q_2(u)  - d_3(u) d_3(u-1) Q_2(u+1) =0\,,
%   \end{aligned}
% \end{equation}
similar to Baxter's TQ-equations for the transfer matrix of the eight-vertex
model \cite{Baxt72a}.
As a consequence of the commutativity of the transfer matrices and the
$Q$-operators among each other for different arguments, analogous third order
difference equations holds for their corresponding eigenvalues.  Using the
analytical properties of the transfer matrices these eigenvalues can be
computed reproducing the result obtained from the ABA \cite{KuRe82}.

Note that the actual solution of the spectral problem for the transfer
matrices by means the Bethe ansatz methods introduced so far is not possible
for all integrable lattice models: the ABA relies on the existence of a
suitable (highest or lowest weight) reference state which does not always
exist, e.g.\ for models with boundary conditions breaking all possible $U(1)$
symmetries.  Similarly, the functional approach based on the TQ-equations
(\ref{vv-TQ}) requires a sufficiently simple (e.g.\ polynomial)
parametrization of the eigenvalues of the $Q$-operators.  Neither of these
requirements is met, e.g., for spin chains with open boundary conditions
subject to non-diagonal boundary fields.  For models with
$U_q[su(2)]$-symmetry this issue has been addressed recently using separation
of variables (SoV) and through the derivation of inversion identities
satisfied by the transfer matrices for certain arguments, both leading to
certain generalizations of the TQ-equations
\cite{Nicc12,CYSW13,KiMN14,FrKa14}.  An added value of the formulation of the
spectral problem within the SoV approach is that it provides a basis for the
proof that the solution is complete \cite{Skly92,FrSW08,FaKN14,KiMN14}.

Application of Sklyanin's SoV approach to integrable $SU(3)$ models leads to
equations similar to the TQ-equations but with $Q_{1,2}$-eigenvalues being
functions on the discrete set of common eigenvalues of the separated
coordinates \cite{Skly93}.  
By explicit construction for small systems we find that,
% In fact, 
similar as in the $SU(2)$ case, this set is contained in the integer spaced
lattice of $u$-values enclosed by the singular points of the difference
equations (\ref{vv-TQ}).
For the model with local spins carrying the fundamental representation, i.e.\
(\ref{vv-ddd}), the spectral parameter $u$ takes values from $\{ u_k,
u_k+1,u_k+2\}_{k=1}^L$.
Eliminating the corresponding amplitudes $Q_{1,2}(u)$ one finds that
\begin{equation}
  \label{vv-uzer}
  U(u_k+1) = 0\,, %\qquad k=1,\ldots,L\,,
\end{equation}
and arrives at the following set of inversion identities for the transfer
matrices
\begin{equation} 
  \label{vv-set_invid}
  \begin{aligned}
   & T(u_k) \, T(u_k -1) = U(u_k) \, , \\
    & T(u_k) \, U(u_k -1) = \Delta(u_k) \, , \\
    & U(u_k) \, U(u_k-1) = \Delta(u_k) \, T(u_k -1) \, ,
  \end{aligned}
\end{equation}
for $k=1,\ldots,L$.  Using the projection property (\ref{vv-proj}) of the
$R$-matrix and the Yang-Baxter relations satisfied by the transfer matrices
$T(u)$ and $U(u)$ these product identities have been derived before by Cao
\emph{et al.} \cite{CYSW14a} (note that the third identity can be obtained
from the two other ones).
These equations, together with the analytical properties
of the transfer matrix are sufficient to compute their eigenvalues.

We end this appendix by noting that -- while we have considered the
$SU(3)$-invariant rational vertex model -- it is straightforward to extend the
discussion to the anisotropic ($q$-deformed) model with trigonometric
dependence of the vertex weights on the spectral parameter
\cite{Bela81,BaVV81,PeSc81,Jimb86a}.

%\bibliography{base,bound,books,frahm}

\end{document}